\documentclass[letterpaper]{JHEP3}
\usepackage{epsfig}

\newcommand{\roughly}[1]{\mathrel{\raise.3ex\hbox{$#1$\kern-0.85em
\lower1ex\hbox{$\sim$}}}}

\newcommand{\lsim}{\roughly<}

\def\pd{\partial}

\def\cA{{\cal A}}

\def\cQ{{\cal Q}}

\def\cF{{\cal F}}

\def\cL{{\cal L}}

\def\cO{{\cal O}}
\def\cR{{\cal R}}
\def\cT{{\cal T}}
\def\cU{{\cal U}}

\def\exd{{\hbox{d}}}

\def\ba{\begin{eqnarray}}
\def\ea{\end{eqnarray}}
\def\be{\begin{equation}}
\def\ee{\end{equation}}

\def\ssJ{{\scriptscriptstyle J}}
\def\ssM{{\scriptscriptstyle M}}
\def\ssN{{\scriptscriptstyle N}}
\def\ssP{{\scriptscriptstyle P}}
\def\ssQ{{\scriptscriptstyle Q}}

\def\ssS{{\scriptscriptstyle S}}
\def\ssT{{\scriptscriptstyle T}}

\def\KK{{\scriptscriptstyle KK}}
\def\QCD{{\scriptscriptstyle QCD}}

\def\nn{\nonumber}

\def\({\left(}
\def\){\right)}

\def\pphi{v}
\def\vaceng{\varrho_\eff}

\def\pref#1{(\ref{#1})}
\def\eff{{\rm eff}}

\title{Bulk Axions, Brane Back-reaction and Fluxes}

\author{ C.P.~Burgess${}^{1,2}$ and L.~van Nierop${}^1$\\

${}^1$Department of Physics \& Astronomy, McMaster University\\ \qquad 1280 Main Street West, Hamilton ON, Canada.\\

${}^2$Perimeter Institute for Theoretical Physics\\
\qquad 31 Caroline Street North, Waterloo ON, Canada.\\
}

\date{}

\abstract { Extra-dimensional models can involve bulk pseudo-Goldstone bosons (pGBs) whose shift symmetry is explicitly broken only by physics localized on branes. Reliable calculation of their low-energy potential is often difficult because it requires an understanding of the dynamics that stabilizes the geometry of the extra dimensions. Rugby ball solutions provide simple examples of extra-dimensional configurations for which two compact extra dimensions are stabilized in the presence of only positive-tension brane sources. The effects of brane back-reaction can be computed explicitly for these systems, allowing the calculation of the shape of the low-energy pGB potential, $V_{4D}(\varphi)$, as a function of the perturbing brane properties, as well as the response of both the extra dimensional and on-brane geometries to this stabilization. If the $\varphi$-dependence is a small part of the total brane tension a very general analysis is possible, permitting an exploration of how the system responds to frustration when the two branes disagree on what the proper scalar vacuum should be. We show how the low-energy potential is given by the sum of brane tensions (in agreement with common lore) when only the brane tensions couple to $\varphi$. We also show how a direct brane coupling to the flux stabilizing the extra dimensions corrects this result in a way that does {\em not} simply amount to the contribution of the flux to the brane tensions. The mass of the low-energy pseudo-Goldstone mode is of order $m_a \sim (\mu/F)^2 m_\KK$ (where $\mu$ is the energy scale associated with the brane symmetry breaking and $F < M_p$ is the extra-dimensional axion decay constant). In principle this can be larger or smaller than the Kaluza-Klein scale, $m_\KK$, but when it is larger axion properties cannot be computed purely within a 4D approximation (as they usually are). We briefly describe several potential applications, including a brane realization of `natural inflation,' and a dynamical mechanism for suppressing the couplings of the pGB to matter localized on the branes. Since the scalar can be light enough to be relevant to precision tests of gravity (in a technically natural way) this mechanism can be relevant to evading phenomenological bounds. }

\begin{document}

\section{Introduction and Summary}

Brane-world models --- where known particles are localized on surfaces within extra dimensions --- have proven to be fruitful places to seek novel kinds of low-energy physics. As studies over the past decade show, their low-energy physics can be novel (relative to 4D models, say) because of several different mechanisms:

\medskip\noindent{\em Brane vs Bulk Kinematics:} Because not all particles are trapped on the branes, different species experience the kinematics of different dimensions. This observation is what allows the existence of unusually large extra dimensions and low gravity scales \cite{ADD,braneworld}.

\medskip\noindent{\em Brane Back-reaction:} Because branes can be localized within the extra dimensions, which need not be homogeneous, physical properties can vary from place to place within the extra dimensions. Such position dependence is generic once the back-reaction of localized branes onto their geometry is included \cite{RS}.

\medskip\noindent{\em Dimensional stabilization:} Because low-energy degrees of freedom can be lighter than the Kaluza-Klein (KK) scale, predictions for low-energy dynamics generically require an equally complete understanding of whatever physics stabilizes the extra dimensions \cite{GW}.

\medskip\noindent
These last two points in particular considerably complicate discussions of cosmology within a brane-world framework \cite{BWcosmo}. What appear to be shallow directions for the scalar potential on a brane for fixed bulk geometry can turn into much steeper directions once the bulk geometry is allowed to move.

With a few exceptions, the interplay between brane back-reaction and the physics stabilizing the bulk remains relatively poorly explored. The main exception is the case of codimension-1 branes moving within one extra dimension, as for Randall-Sundrum (RS) models \cite{RS}. Yet one wonders how representative codimension-1 systems are of the more generic situation having higher codimension, for which much less is known about brane back-reaction. Although some results exist for the back-reaction of branes on ten-dimensional geometries in string theory \cite{KKLMMT}, it is often not possible to be as explicit about the form of the extra-dimensional geometry and its detailed interplay with brane back-reaction.

In this paper we compute the low-energy potential for a class of codimension-2 brane models for which explicit compactifications involving both bulk stabilization and brane back-reaction are known. We focus in particular on the simplest of brane/flux compactifications: (nonsupersymmetric) 6D Einstein-Maxwell theory with the extra dimensions stabilized through a competition between a background Maxwell flux and a bulk cosmological constant \cite{6dnonsusy,conical,SalSez,Towards,6dsusy}. To this system we couple a bulk Goldstone boson (axion), $\phi$, whose shift symmetry, $\phi \to \phi + $(constant), is explicitly broken by its couplings to the two branes that source the bulk through interactions of the schematic form
\be
\label{braneaction1}
 S_b = \int_{\Sigma_b}  \Bigl( \tau_b \, \omega + \Phi_b \, {}^\star \cF  \Bigr) \,,
\ee
where the integration is over the 4D brane world-sheet, $\Sigma_b$, whose volume form is denoted $\omega$ so $\tau_b(\phi)$ represents a $\phi$-dependent brane tension. $\cF_{\ssM\ssN}$ denotes the 6D Maxwell field strength --- for which ${}^\star \cF$ is the 6D Hodge dual --- and the $\phi$-dependent coupling $\Phi_b(\phi)$ can be interpreted as the amount of Maxwell flux carried by the brane in question.

By computing the back-reaction of the branes onto the bulk fields we obtain the low-energy potential for the resulting would-be Goldstone zero mode, $\varphi$, that becomes a pseudoGoldstone boson (pGB) \cite{pGB} in the low-energy 4D effective theory. The branes back-react onto the bulk fields by changing their boundary conditions, through a codimension-2 generalization \cite{cod2matching,BBvN,otheruvcaps} of the more familiar codimension-1 Israel junction conditions \cite{IJC}.

The resulting bulk field equations subject to the brane boundary conditions can be solved explicitly in some generality if the axion-dependence of the brane tension is regarded as a small change to a background, axion-independent value. In this limit the would-be zero mode is stabilized to a fixed value, $\varphi = \varphi_\star$, which we compute in two separate ways: first by explicitly solving the linearized field equations of the full 6D theory; and second by minimizing the dimensionally reduced axion potential in the 4D low-energy effective theory. Both methods agree, and the generality of our result allows us to follow how the stabilized value and its energy density vary as a function of the axion couplings to the two branes. In particular we see what happens when the branes differ in the value for the axion that they prefer. As a by-product we also compute how the geometry of the extra dimensions changes due to the presence of the axion-brane couplings.

The calculation reveals the following generic features
\begin{enumerate}
\item In the absence of brane fluxes --- {\em i.e.} $\Phi_b = 0$ --- the low-energy 4D potential is very generally simply given by the sum of tensions, summed over the branes present,
    \be \label{introVeff}
     V_\eff(\varphi) = \sum_b \tau_b[ \phi_b(\varphi) ] \,,
    \ee
    where $\phi_b(\varphi)$ denotes the value taken by the (suitably renormalized) 6D scalar field at the corresponding brane position, regarded as a function of the zero mode $\varphi$. This agrees with the probe-brane approximation (which ignores brane back-reaction) since for brane tensions the contribution of the back-reaction first arises at second order. In particular the stabilized value, $\varphi = \varphi_\star$, satisfies
    \be \label{introphistar}
    \sum_b \left( \frac{\partial \tau_b}{\partial \phi} \right)_{\varphi = \varphi_\star} = 0 \,.
    \ee
    Because the quantity $(\partial \tau_b/\partial \phi)_{\varphi_\star}$ governs the coupling of the lightest mode, $\varphi$, to matter localized on the brane, these couplings tend naturally to turn themselves off for small fluctuations of $\varphi$ about its ground state. This could provide a phenomenologically useful mechanism for naturally decoupling light bulk scalars from brane matter, along the lines of similar earlier proposals \cite{natdecouple}.

\item By contrast, nonzero brane fluxes contribute at linear order in two ways, that are similar in size. The first arises because nonzero background fluxes are required to stabilize the bulk geometry, ${}^\star \cF = -\cQ \, \omega$ (with, as before, $\omega$ the volume form). Because of this the $\Phi_b {}^\star \cF$ term modifies the value of the brane action, giving an `effective tension'
    \be \label{introTb}
     T_b (\phi) = \tau_b(\phi) - \cQ \, \Phi_b(\phi) \,.
    \ee
    The second contribution arises because quantization of total flux requires the amount of bulk flux to change in response to the presence of flux on the brane, leading to an additional energy cost over and above that measured by the difference $T_b - \tau_b$. Although the complete expression for $V_\eff(\varphi)$ that results involves an integration of $\Phi_b$ with respect to $\varphi$, the predictions for $\varphi_\star$ and $\vaceng := V_\eff(\varphi_\star)$ turn out to be relatively simple. The prediction for $\varphi_\star$ is again given by eq.~\pref{introphistar} --- with no contribution from $\Phi_b$ --- while the prediction for $\vaceng$ becomes
    \be
     \vaceng = \sum_b \Bigl\{ T_b[\phi_b(\varphi_\star)] - \cQ \, \Phi_b[\phi_b(\varphi_\star)]
     \Bigr\} = \sum_b \Bigl\{ \tau_b[ \phi_b(\varphi_\star) ] - 2 \cQ \, \Phi_b[ \phi_b(\varphi_\star)]
     \Bigr\} \,.
    \ee
    We see in this way that the brane flux `contributes twice' to the vacuum energy at low energies. In particular, $\vaceng = 0$ for branes satisfying $\tau_b[\phi_b(\varphi_\star)] = 2\cQ \, \Phi_b[\phi_b(\varphi_\star)]$, and so $T_b[\phi_b(\varphi_\star)] = \cQ \, \Phi_b[ \phi_b(\varphi_\star)]$.

\item Assuming a 6D kinetic energy of the form $F^4 \partial^\ssM \phi \, \partial_\ssM \phi$, in order of magnitude the mass of the light 4D would-be zero mode, $\varphi$, predicted by the low-energy potential is of order $m_\varphi \sim (\mu/F)^2 m_{\KK} f(\varphi_\star)$, where $m_\KK$ denotes the KK mass scale and the $\phi$-dependent part of the brane tension is assumed to be of order $\mu^4$. In most cases of interest $f(\varphi_\star)$ is given by $\sum_b \partial^2 \tau_b/\partial \phi^2$ evaluated at $\varphi_\star$. When $\mu \ll F$ this mode satisfies $m_\varphi \ll m_\KK$, allowing its properties to be described in the effective 4D theory. (Unlike for a purely 4D theory it makes sense to call $\phi$ a pseudo-Goldstone field even if $\mu > F$, since its shift symmetry is everywhere unbroken in the extra dimensions except at the positions of the branes. However, if $\mu > F$ then the mass of the would-be zero mode becomes comparable with other KK modes, precluding calculating its properties within the 4D theory.\footnote{Exceptions to this can arise if the scalar is self-localized at the brane \cite{selfloc}, but this usually requires a bulk scalar potential $U(\phi)$ that is forbidden in the current examples by the assumed shift symmetry, $\phi \to \phi + c$.})
\end{enumerate}

The low-energy 4D theory obtained from models with bulk axions generically includes a light scalar field, whose small mass is technically natural because of the weakly broken shift symmetry. What the brane construction potentially provides is a UV completion that can explain why the masses and couplings are small in the first place. This could prove useful for a variety of low-energy applications, such as to extra-dimensional inflationary models, some of which we briefly describe below while examining specific examples of our general expressions.

We organize our detailed discussion as follows. Our main results are presented in the next section, which starts in \S2.1 by setting out the field equations describing the system of interest. These are then solved by finding solutions that are perturbatively close to simple, well-known rugby-ball solutions involving two branes interacting with a spherical 2D bulk. \S2.2 does so first for the simpler case where the branes do not couple to the bulk scalar, with the generalization to scalar-brane coupling following in \S2.3. \S3 then explores the features of these general solutions by examining in detail several simple illustrative special cases. These include situations where the two branes agree on the value at which the field $\varphi$ stabilizes, and situations where they do not. \S3 also considers several special cases of potential phenomenological interest for axion and inflationary applications.

\section{The bulk-brane system}

This section defines the system of interest, which we take to be the simplest theory containing both codimension-two sources (with positive tension) and a flux mechanism for stabilizing the size of the extra dimensions. This suggests taking the bulk theory to be 6D Einstein-Maxwell gravity coupled to the Goldstone boson (axion) field.

\subsection{Field equations and background solutions}

Classical brane-bulk dynamics is defined by solving the bulk field equations, subject to the boundary conditions imposed by matching conditions at each brane.

\subsubsection*{Bulk field equations}

The bulk action of interest is\footnote{We use a `mostly plus' metric and Weinberg's curvature conventions \cite{Wbg} (that differ from those of MTW \cite{MTW} only by an overall sign in the definition of the Riemann tensor).}
\be
\label{bulk-brane system: bulkaction}
 S_{\rm bulk} = - \int \exd^6x \, \sqrt{-g} \; \left\{ \frac12 \,
 g^{\ssM \ssN} \left( \frac{1}{\kappa^2} \, \cR_{\ssM \ssN}
 + \frac{1}{\kappa_a^2} \, \pd_\ssM \phi \, \pd_\ssN \phi \right)
 + \frac14 \, \cF_{\ssM\ssN} \cF^{\ssM\ssN} + \Lambda \right\} \,,
\ee
where $\cR_{\ssM\ssN}$ is the Ricci tensor constructed from the 6D metric $g_{\ssM \ssN}$, $\phi$ is the axion field and $\cF = \exd \cA$ is the field strength for the Maxwell potential $\cA_\ssM$. The dimensionful parameters of the problem are the 6D gravitational coupling, $\kappa = 1/M_g^2$, the bulk axion decay constant, $\kappa_a = 1/F^2$, and the bulk cosmological constant, $\Lambda$ (whose value is tuned to ensure the unperturbed solution is flat in the on-brane directions). Although $\kappa_a$ can be absorbed into the normalization of $\phi$, we do not do so because this changes the form of the brane couplings to $\phi$.

The field equations obtained from this action are the (trace-reversed) Einstein equation
\be
 \cR_{\ssM\ssN} + \lambda^2 \, \partial_\ssM \phi \, \partial_\ssN \phi
 + \kappa^2  \cF_{\ssM \ssP} {\cF_\ssN}^\ssP
  - \left[ \frac{\kappa^2}{8} \, \cF_{\ssP\ssQ} \cF^{\ssP \ssQ}
 - \frac{\kappa^2 \Lambda}{2}  \right] g_{\ssM\ssN} = 0 \,,
\ee
where $\lambda := \kappa/\kappa_a = F/M_g$. The Maxwell equation is
\be
 \sqrt{-g} \; \nabla_\ssM \cF^{\ssM \ssN} = \partial_\ssM \Bigl(
 \sqrt{-g} \; \cF^{\ssM \ssN} \Bigr) =  0 \,,
\ee
and
\be
 \sqrt{-g} \; \Box \phi = \pd_\ssM \left( \sqrt{-g} \; \pd^\ssM\phi \right) = 0 \,,
\ee
is the axion equation.

\subsubsection*{Rugby-ball solutions}

We consider geometries that are maximally symmetric in the 4 on-brane directions and axially symmetric in the two extra dimensions. The corresponding {\em ansatz} for the metric, scalar and Maxwell fields is
\be
\label{bulk-brane system: ansatz}
 \exd s^2 = \exd\rho^2 + e^{2B} \exd\theta^2 + e^{2W} \,
 \hat g_{\mu\nu} \exd x^\mu \exd x^\nu \,,
\ee
and
\be
 \cF_{\rho\theta} = \cA_\theta' \,,
\ee
where $\hat g_{\mu\nu}$ is an $x^\mu$-dependent maximally symmetric geometry and the functions $B$, $W$, $\phi$ and $\cA_\theta$ depend only on $\rho$. Primes denote differentiation with respect to this coordinate.

Subject to this {\em ansatz} the bulk field equations reduce to
\ba
\label{bulk-brane system: EOM}
 \left( e^{B+4W} \, \phi' \right)' &=& 0 \qquad (\phi) \nn\\
 \left( e^{-B+4W} \cA_{\theta}' \right)' &=& 0 \qquad (\cA_\theta) \nn\\
 4 \Bigl[ W'' + (W')^2 \Bigr] + B'' + (B')^2 + \lambda^2 (\phi')^2
 + \frac{3\kappa^2}{4} \, e^{-2B} \left( \cA_\theta' \right)^2
 + \frac{\kappa^2 \Lambda}{2} &=& 0\qquad (\rho\rho) \nn\\
 B'' + (B')^2 + 4W'B' + \frac{3\kappa^2}{4} \, e^{-2B} \left(
 \cA_{\theta}' \right)^2 + \frac{\kappa^2 \Lambda}{2} &=& 0
 \qquad (\theta\theta) \nn\\
 \frac14 \, e^{-2W} \hat \cR + W'' + 4(W')^2 + W'B'
 - \frac{\kappa^2}{4} \, e^{-2B} \left( \cA_{\theta}' \right)^2
 +\frac{\kappa^2 \Lambda}{2} &=& 0 \qquad (\mu\nu) \,,
\ea
where $\hat \cR$ is the curvature scalar built from the maximally symmetric metric $\hat g_{\mu\nu}$.
The first two of these immediately integrate to give
\be \label{phiAfirstintegrals}
  e^{B+4W} \, \phi' = \varphi_1 \quad \hbox{and} \quad
  e^{-B+4W} \cA_{\theta}' = \cQ \,,
\ee
where $\varphi_1$ and $\cQ$ are integration constants.

When $\varphi_1 = 0$ the full set of equations admits a particularly simple solution of the rugby-ball form \cite{6dnonsusy}
\ba
\label{bulk-brane system: rugby ball}
 \exd s^2 &=& \exd \rho^2 + \alpha^2 L^2 \sin^2 \left( \frac{\rho}{L}
 \right) \exd \theta^2 + \hat g_{\mu\nu} \, \exd x^\mu \exd x^\nu \nn\\
 \cF_{\rho\theta} &=& \cQ \alpha L \, \sin\left( \frac{\rho}{L} \right) \,,
\ea
with constant $\phi = \varphi_0$ and $W = 0$. The equations of motion imply the following relation amongst the integration constants:
\be
 \frac2{L^2} = \kappa^2 \left( \frac{3\cQ^2}{2} + \Lambda \right) \,,
\ee
as well as fixing the 4D curvature
\be
 \hat{\cR} := \hat g^{\mu\nu} \hat \cR_{\mu\nu}
 = \kappa^2 \left( \cQ^2 - 2 \Lambda \right) \,.
\ee

A final constraint relating parameters comes from flux quantization, due to the spherical topology of the extra dimensions. As usually framed, this implies
\be
 \frac ng = 2 \alpha L^2 \cQ \,,
\ee
where $g$ is the gauge coupling of the Maxwell field and $n$ is an arbitrary integer. However this expression assumes the absence of any flux localized on the source branes themselves \cite{Towards}. In the presence of brane-localized flux (more about this below and in Appendix \ref{App:fluxconditionws}) the flux-quantization condition instead becomes
\be
 \frac ng=2\alpha L^2\cQ + \sum_{ b} \frac{\Phi_b}{2\pi} \,,
\ee
where the sum is over all of the branes present, each of which carries the localized flux, $\Phi_b$.

Since our interest is in background solutions with flat geometries, $\hat g_{\mu\nu} = \eta_{\mu\nu}$, we further choose $\Lambda$ so that $\hat \cR = 0$:
\be
 \Lambda = \frac{\cQ^2}{2}
 \quad \hbox{and so} \quad
 \kappa^2 L^2 \cQ^2 = 1 \,.
\ee
With this choice all geometrical properties, like $L$ and $\cQ$, can be regarded as functions of the integration constant $\alpha$ together with the integer $n$ and the lagrangian parameters $\kappa$ and $g$ (see Appendix \ref{App:rugbyballresponse} for details).

The potential singularity in the geometry where $g_{\theta\theta} = e^B$ vanishes is just a coordinate artefact when $\alpha = 1$, in which case the two compact dimensions define a sphere. When $\alpha \ne 1$ the background has a conical singularity at $\rho = \rho_\ssN := 0$ and $\rho = \rho_\ssS := \pi L$. This is interpreted as describing the back-reaction of two codimension-two source branes located at these positions, having equal tensions, $T$ (which includes the energy associated with the brane flux, see eqn.~\pref{bulk-brane system: braneaction}. Matching at the branes (see below) implies this tension is related to the deficit angle  by
\be \label{deficitangle}
 1-\alpha = \frac{\kappa^2 T}{2\pi} \,,
\ee
similar to the relation between tension and deficit angle for a cosmic string \cite{conicalmatching}.

Finally, notice that the value $\phi = \varphi_0$ is not determined by any of the equations of motion, due to the symmetry $\phi \to \phi + \hbox{constant}$. The parameter $\varphi_0$ labels a flat direction in the low-energy potential, that can be lifted if the coupling of $\phi$ to the branes breaks this symmetry (such as by allowing the tensions $T$ to depend on $\phi$). A primary goal of the next few sections is to identify the effective potential for this low-lying mode below the KK scale, to determine how the vacuum value after symmetry breaking, $\varphi_*$, is related to the couplings on the branes.

\subsubsection*{Brane matching conditions}

As brane sources we use the most general form (involving the fewest derivatives) for a 4D brane action located at positions $\rho_\ssN$ and $\rho_\ssS$ \cite{Towards}
\ba
\label{bulk-brane system: braneaction}
 S_\mathrm{branes} &=& - \sum_{b=\ssN,\ssS} \int \exd^4x \,\sqrt{- g_4} \;  \left[
 \tau_b - \frac{\Phi_b}{2} \, \epsilon^{mn} \cF_{mn} \right] \nn\\
 &=& - \sum_{b=\ssN,\ssS} \int \exd^4x \,\sqrt{- \hat g} \; e^{4W} \left[
 \tau_b - \frac{\Phi_b}{2} \, \epsilon^{mn} \cF_{mn} \right] \,,
\ea
where $\epsilon^{\rho \theta} = 1/\sqrt{g_2} = e^{-B}$ transforms as a tensor, rather than a tensor density, in the two transverse dimensions. The parameter $\tau_b$ represents the tension of the brane, which can depend on all of $\phi$, $W$ and $g_{\theta\theta}$ without breaking the condition of maximal symmetry in the on-brane directions. As is shown below, the parameter $\Phi_b$ similarly denotes the magnetic charge (or flux) carried by the source branes (which could also depend on $\phi$, $W$ and $g_{\theta\theta}$).

The presence of such branes imposes a set of boundary conditions on the derivatives of the bulk fields in the near-brane limits, given by\footnote{Notice that we normalize the quantities $\cT_b$ and $\cU_b$ without including a factor of $e^{4W}$ used in ref.~\cite{BBvN}.}
\begin{eqnarray} \label{matching}
 \Bigl[ e^{B} {\phi}' \Bigr]_{\rho_b}
 &=& \frac{\partial
 \cT_b}{\partial \phi} \quad \hbox{with} \quad
 \cT_b := \frac{\kappa^2 T_b}{2\pi}  \nn\\
 \Bigl[ e^{B} W' \Bigr]_{\rho_b}
 &=&  \cU_b \quad \hbox{with} \quad
 \cU_b :=  \frac{\kappa^2}{4\pi} \left(
 \frac{\partial T_b}{\partial g_{\theta\theta}} \right)   \\
 \hbox{and } \quad
 \Bigl[ e^B B'-1 \Bigr]_{\rho_b}
 &=& - \Bigl[ \cT_b +  3\,\cU_b
 \Bigr] \,,\nn
\end{eqnarray}
where $T_b$ is defined as the total lagrangian density of the source,
\be
 T_b = \tau_b - \, \Phi_b \, e^{-B} F_{\rho\theta} \,.
\ee
The Bianchi identities ensure that only two of eqs.~\pref{matching} are independent of one another, and as a consequence the quantities $\cU_b$ and $\cT_b$ are also not independent. They are subject to the constraint:
\begin{equation}
\label{curvature constraint}
  4\cU_b \Bigl[ 2  - 2\cT_b - 3 \, \cU_b \Bigr]
  - (\cT_b')^2
  = 0 \,,
\end{equation}
where $\cT_b' = \partial \cT_b / \partial \phi$. Notice that for the rugby ball solutions $\cT_b' = \cU_b = 0$ and so eqs.~\pref{matching} degenerate down to eq.~\pref{deficitangle}.

As shown in Appendix \ref{App:fluxconditionws}, the corresponding boundary condition for the Maxwell field implies that the integral of eq.~\pref{phiAfirstintegrals} for $\cA_\theta(\rho)$ for a patch containing each source brane is \cite{Towards}
\ba
 \cA_\theta(\rho) &=& \frac{\Phi_\ssN}{2\pi} + \cQ \int_{\rho_\ssN}^\rho
 \exd \hat \rho  \; e^{B-4W} \quad\;\; \hbox{Northern hemisphere} \nn\\
 &=& - \frac{\Phi_\ssS}{2\pi} + \cQ \int_{\rho_\ssS}^\rho \exd \hat \rho
 \; e^{B-4W}  \quad \hbox{Southern hemisphere} \,,
\ea
and the signs are dictated by the observation that increasing $\rho$ points away from (towards) the North (South) pole, together with the requirement that the two patches share the same orientation. Requiring these to differ by a gauge transformation, $g^{-1} \partial_\theta \Omega$, on regions of overlap implies the flux-quantization condition
\be \label{fluxquantzn}
 \frac{n}{g} = \frac{\Phi_{\rm tot}}{2\pi}
 + \cQ \int_{\rho_\ssN}^{\rho_\ssS} \exd \rho \; e^{B - 4W} \,,
\ee
where $n$ is an integer, $g$ is the gauge coupling and $\Phi_{\rm tot} = \Phi_\ssN + \Phi_\ssS$. It is this expression that identifies $\Phi_b$ as the fraction of the total Maxwell flux carried by each brane.

\subsection{Perturbations I: the Einstein-Maxwell case}

Next consider starting with a rugby-ball solution and independently perturbing each of the two brane tensions, $\tau_b = \tau + \delta \tau_b$, and brane-localized fluxes, $\Phi_b = \Phi+\delta \Phi_b$, implying a similar expansion for the total brane action, $T_b = \tau_b -e^B \cF_{\rho\theta}  \Phi_b = \tau_b - \cQ \, \Phi_b$.  This section starts simply and assumes both $\delta \tau_b$ and $\delta \Phi_b$ are independent of $\phi$, with the resulting insights used to inform the next section's discussion of the more general case. The goal is to compute explicitly how the bulk fields respond to the perturbation, allowing a detailed examination of how the extra dimensions flex as their source branes change. In general, because the perturbed branes are different from one another, the scalar field acquires a nontrivial profile, $\phi = \phi(\rho)$, and the resulting geometry warps nontrivially, $W = W(\rho)$.

\subsubsection*{Linearized solutions}

Because the brane perturbations are independent of $\phi$, $\partial T_b/\partial \phi = 0$ and so there is no change to the $\phi$ boundary conditions. Consequently the unperturbed solution, $\phi = \varphi_0$, remains a solution. The scalar then drops out of the problem and the calculation involves only the Einstein-Maxwell system. Writing $e^B = e^{B_0}[1 + \delta B(\rho)]$ and $W = \delta W(\rho)$ --- with $B_0$ given by the rugby-ball solution described by parameters $\cQ$, $L$ and $\alpha$ ---  we linearize the field equations in $\delta B$ and $\delta W$.

The combination of the Einstein equations $(\rho\rho)-(\theta\theta)$ linearizes to
\be
 \delta W'' - B_0' \, \delta W' = \delta W''
 - \frac{\delta W'}{L} \cot \left( \frac{\rho}{L} \right)
 = 0
\ee
which has as its solution
\be \label{Wsolnsimple}
 \delta W = W_1 \cos \left( \frac{\rho}{L} \right) \,,
\ee
where we absorb an additive integration constant, $W_0$, into a re-scaling of the four on-brane coordinates, $x^\mu$.

The perturbed gauge field again satisfies eq.~\pref{phiAfirstintegrals}, and so
\be
 \delta \cA_\theta' = \Bigl[ \delta \cQ + \cQ \left( \delta B - 4 \delta W \right)
 \Bigr] \alpha L \sin \left( \frac{\rho}{L} \right) \,.
\ee
With this and eq.~\pref{Wsolnsimple} the $\theta\theta$ Einstein equation linearizes to
\be
 \frac{[\delta B' \sin^2 \left( {\rho}/{L} \right) ]'}{\sin^2 (\rho/L)}
 = \frac{10 \, W_1}{L^2} \; \cos\left( \frac{\rho}{L} \right)
 - \frac{3}{2L^2} \left( \frac{\delta Q}{Q} \right) \,,
\ee
whose integral is
\be
 \delta B = \frac34 \left( \frac{\delta\cQ}\cQ \right)
 \frac\rho L \, \cot \left( \frac{\rho}{L} \right)
 -\frac{10W_1}{3} \, \cos \left( \frac{\rho}{L} \right)
 - B_1 \cot \left( \frac{\rho}{L} \right) + \delta B_0 \,.
\ee
Notice that the integration constant $B_1$ here is pure gauge, corresponding to an infinitesimal shift in the radial coordinate $\rho \to \rho + c$. We can fix this freedom by demanding that $\rho_\ssN = 0$, and so $e^{B} \to 0$ as $\rho \to 0$. Since $e^B = \alpha L (1 + \delta B) \sin (\rho/L) \to - B_1$ at $\rho \to 0$, this implies $B_1 = 0$.

The linearized flux quantization condition, eq.~\pref{fluxquantzn} is
\be
 \frac{\delta\cQ}\cQ + \frac12 \int_0^{\pi L} \frac{\exd\rho}L
 \sin \left( \frac\rho L \right) (\delta B - 4 \delta W)
 + \frac{\kappa^2 \cQ}{4\pi\alpha} \left(\delta\Phi_\ssN + \delta\Phi_\ssS \right) =0 \,,
\ee
which uses the background relation $\kappa L \cQ = 1$ to rewrite $\delta\Phi_b/L^2 \cQ = \kappa^2 \cQ \delta\Phi_b$. Solving this for $\delta\cQ/\cQ$ gives
\be
 \frac{\delta\cQ}\cQ = -4 \delta B_0 -\frac{\kappa^2 \cQ}{\pi\alpha} \left(\delta\Phi_\ssN+\delta\Phi_\ssS\right) \,.
\ee

In summary, once coordinate conditions are used to eliminate $W_0$ and $B_1$, solutions to the bulk equations for $\delta W$, $\delta B$ and $\delta \cA_\theta$ involve three integration constants --- $W_1$, $\delta B_0$ and $\delta\cQ/\cQ$ --- among which flux quantization imposes one relation. The physical interpretation of the two remaining parameters is seen by connecting them to two physical quantities. One of these can be taken as the warping difference between the two branes,
\be
 \delta W_\ssN - \delta W_\ssS = 2 W_1 \,,
\ee
which controls the relative redshift of energies on the two branes. The other can be chosen as the change in proper distance, $\rho_\ssS - \rho_\ssN = \pi (L + \delta L)$, between the two branes, where $\rho_\ssN = 0$ and $\rho_\ssS$ are defined as the places where $e^B \to 0$. Comparing
\be
 \lim_{\rho \to \pi L} e^B \simeq - \frac{3\pi \alpha L}4
 \left( \frac {\delta \cQ} \cQ \right) \,,
\ee
with the Taylor expansion $e^B (\pi L) \simeq (e^{B_0})'_{\rho_\ssS} (-\pi \delta L) = + \pi \alpha \delta L$ gives
\be
 \frac{\delta L} L \simeq - \frac34 \left( \frac {\delta \cQ} \cQ \right) \,.
\ee

\subsubsection*{Matching to brane tensions}

All that remains is to eliminate the integration constants $\delta B_0$ and $W_1$ in terms of the brane perturbations using the linearized brane matching conditions. In the present instance only the last of eqs.~\pref{matching} is nontrivial. Besides imposing the background relation $\alpha=1-\kappa^2T/(2\pi)$, for the linearized perturbations this condition implies
\be \label{Bwithcurrent}
 \delta \Bigl( e^{B} \Bigr)_{\rho_b}' =  -\frac{\kappa^2 \delta T_b}{2\pi} \,,
\ee
where $\delta T_b = \delta( \tau_b - \Phi_b \cQ e^{-4W_b}) \simeq \delta \tau_b - \delta \Phi_b \cQ - \Phi\cQ (\delta\cQ/\cQ - 4 \delta W_b)$. Evaluating this at $\rho = \rho_\ssN = 0$ and\footnote{To leading order $\delta\rho_\ssS$ does not contribute, and is only mentioned for completeness.} $\rho = \rho_\ssS = \pi L + \delta \rho_\ssS$, and keeping in mind that it is $-\rho$ that is the outward direction for the south brane, gives
\ba \label{WBlineqs}
 \alpha \left[ \delta B_0 - \frac{10 W_1}{3} + \frac34 \left( \frac{\delta
 \cQ}{\cQ} \right) \right] &=&
  \alpha \left[-2 \delta B_0 - \frac{10 W_1}{3} - \frac{3\kappa^2 \cQ
  \delta\Phi_{\rm tot} }{4\pi \alpha} \right]
  = - \frac{\kappa^2 \delta T_\ssN}{2\pi} \nn\\
 \alpha \left[ \delta B_0 + \frac{10 W_1}{3} + \frac34 \left( \frac{\delta
 \cQ}{\cQ} \right) \right] &=&
 \alpha \left[- 2 \delta B_0 + \frac{10 W_1}{3} - \frac{3\kappa^2 \cQ
 \delta\Phi_{\rm tot} }{4\pi \alpha} \right]
 = - \frac{\kappa^2 \delta T_\ssS}{2\pi} \,.
\ea
When solving these we may approximate $\delta T_b \simeq \delta \tau_b - \cQ \, \delta\Phi_b$, which involves dropping the back-reaction of those terms proportional to $\delta \cQ/\cQ$ and $\delta W_b$ in $\kappa^2 \delta T_b/2\pi$. This neglect is justified because their relative contribution is of order $\kappa^2\cQ\Phi/2\pi$, which must be small to justify our classical treatment of gravity. The solution found within this approximation to eqs.~\pref{WBlineqs} then is
\ba
 \delta B_0 &=& \frac{\kappa^2}{8\pi\alpha} \Bigl[ \delta T_\ssN +
 \delta T_\ssS - 3 \cQ ( \delta\Phi_\ssN + \delta\Phi_\ssS )\Bigr]\nn\\
 W_1 &=& \frac{3\kappa^2}{40\pi \alpha} \Bigl( \delta T_\ssN - \delta T_\ssS
 \Bigr) \,,
\ea
Using the above,
\ba
 \frac{\delta L}{L} = - \frac34 \left( \frac{\delta \cQ}{\cQ} \right)
 &=& \frac{3 \kappa^2}{8\pi \alpha} \Bigl[ (\delta T_\ssN + \delta T_\ssS)
 - \cQ ( \delta\Phi_\ssN + \delta\Phi_\ssS) \Bigr] \nn\\
 &=& \frac{3 \kappa^2}{8\pi \alpha} \Bigl[ (\delta \tau_\ssN + \delta \tau_\ssS)
 - 2\cQ ( \delta\Phi_\ssN + \delta\Phi_\ssS)\Bigr] \,.
\ea

\subsubsection*{On-brane geometry and the view from 4D}

The curvature of the induced geometry on the branes comes from the linearized $(\mu\nu)$ Einstein equation, which --- using $e^{-2B} F_{\rho\theta}^2 = \cQ^2 e^{-8W}$ --- gives
\ba
 \hat \cR &=& -4 \left[ \delta W'' + \frac{\delta W'}{L} \, \cot \left( \frac \rho L \right) \right]
 - \frac{8 \delta W }{L^2} + \frac{2}{L^2} \left( \frac{\delta\cQ}{\cQ} \right)
 = \frac{2}{L^2} \left( \frac{\delta\cQ}\cQ \right) \nn\\
 &=& - \frac{\kappa^2}{\pi\alpha L^2} \Bigl[ (\delta \tau_\ssN + \delta \tau_\ssS)
 - 2\cQ (\delta\Phi_\ssN + \delta\Phi_\ssS) \Bigr] \,.
\ea

{}From the point of view of a 4D observer localized on the brane this curvature would be interpreted as being due to a 4D energy density, $\vaceng$. Since the 4D gravitational coupling, $\kappa^2_4 = 8 \pi G_\ssN$, is related to $\kappa$ by
\be \label{4DkappaDR}
 \frac{1}{\kappa^2_4} = \frac{2\pi \alpha L}{\kappa^2} \int_0^{\pi L} \exd \rho \;
 \sin \left( \frac \rho L \right) = \frac{4 \pi \alpha L^2}{\kappa^2} \,,
\ee
we have
\be
 \vaceng = -\frac{\hat \cR}{4\kappa_4^2} = \delta \tau_\ssN + \delta \tau_\ssS
 -2 \cQ (\delta\Phi_\ssN + \delta\Phi_\ssS) \,.
\ee
Notice that this agrees with the naive expectation $\vaceng = \delta \tau_\ssN + \delta \tau_\ssS$ in the absence of fluxes on the brane. The same is {\em not} true in the presence of brane fluxes, however, since the final result for $\vaceng$ differs from $\delta T_\ssN + \delta T_\ssS =  \delta \tau_\ssN + \delta \tau_\ssS - \cQ(\delta \Phi_\ssN +\delta \Phi_\ssS)$. As the above calculation shows, $\vaceng \propto -\delta \cQ/\cQ$ and so the energy cost of the perturbation arises from the change of flux required by the flux-quantization condition in response to the back-reaction of the branes on the bulk geometry. Since the flux is homogeneous across the extra dimensions, its energy cost is expensive since it scales with the volume. Localizing some of the flux into the branes reduces this extensive energy cost.\footnote{Of course, this possibility of back-reaction competing with brane tensions is already suggested by the complete absence of on-brane curvature in the initial rugby ball solution despite the presence of the initial equal brane tensions, $T$.}

The comparative importance of such back-reaction effects depends on the relative size of the two brane energy scales $\delta \tau_b$ and $\cQ\delta \Phi_b = \delta\Phi_b/\kappa L$. If both $\delta \tau_b$ and $\delta\Phi_b$ are set by the same scale --- {\em i.e.} $\tau_b \sim \Lambda^4$ and $\delta\Phi_b \sim \Lambda$ --- then $\delta \tau_b \sim \cQ \delta\Phi_b$ when $\Lambda \simeq \Lambda_\star := (\kappa L)^{-1/3}$. For $\Lambda$ smaller than this the $\cQ \delta\Phi_b$ term dominates, while $\delta \tau_b$ is the larger of the two when $\Lambda > \Lambda_\star$. (For a similar setup with $n$ transverse dimensions this crossover would occur when $\Lambda \simeq (\kappa L)^{-2/(4+n)}$.) Although $\kappa \Lambda^2$ must be much smaller than one to justify semiclassical methods, for fundamental objects it is comparatively large ({\em e.g.} for $D$-branes $\kappa \Lambda^2$ is of order the string coupling, $g_s \simeq 0.01$ say), and so the tension contribution can therefore dominate. The flux contribution instead can dominate for lower-tension objects.

It is instructive to check this calculation by directly evaluating the low-energy potential through dimensional reduction of the 6D theory in the classical approximation. A general formula for this is computed (including brane back-reaction) in ref.~\cite{BBvN}, and when this is specialized to linear perturbations about a rugby ball it evaluates to
\be
 V_{\rm eff} = 2\pi \int_0^{\pi L} \exd\rho \; e^{B + 4W}
 \left\{ \frac1{2\kappa^2} \left[ \frac{8W'}{L} \cot\left(\frac\rho L\right)
 \right] - \frac14 (\cQ + \delta\cQ)^2 e^{-8W} + \frac\Lambda 2 \right\} \,,
\ee
with the $W'$ term arising from the extra-dimensional curvature and the $(\cQ + \delta \cQ)^2$ term coming from the bulk Maxwell action. In the present instance all terms involving $\delta W$ in this expression turn out to be proportional to $\sin(\rho/L) \cos(\rho/L)$ to linear order in the perturbations, and so integrate to zero and do not contribute to $V_{\rm eff}$. Keeping in mind the background relations $2\Lambda = \cQ^2$ and $\kappa L \cQ = 1$, the result therefore simplifies to
\ba
 V_{\rm eff} &\simeq& 2\pi \int_0^{\pi L} \exd\rho \; e^{B_0}
 \left\{ \left(- \frac{\cQ^2}{4} + \frac\Lambda 2 \right) (1 + \delta B)
  - \frac{\cQ \delta\cQ}{2}  \right\} \nn\\
 &=& - \frac{2\pi\alpha}{\kappa^2} \left( \frac{\delta\cQ}{\cQ} \right)
 = - \frac{\pi \alpha L^2 \hat \cR}{\kappa^2} = \vaceng \,,
\ea
showing the equivalence between the 4D and 6D perspectives.

\subsection{Perturbations II: the Einstein-Maxwell-axion case}

In this section we generalize the previous discussion to consider branes and fluxes that can depend on $\phi$. This allows us to follow how couplings to the brane lift the flat direction associated with the shift symmetry of the bulk theory, and so to see how the scalar zero mode, $\varphi_0$, becomes stabilized at a specific value, $\varphi_0 = \varphi_\star$.

It is instructive to ask how this stabilization happens from the point of view of the full six-dimensional theory. To this end imagine trying to integrate the field equations to obtain the bulk configuration that interpolates between the two branes. Specializing to solutions that are both axially symmetric in the transverse directions and maximally symmetric in the on-brane dimensions we seek bulk profiles as a function only of $\rho$, starting with initial conditions set by matching to the brane at $\rho = \rho_\ssN$ (say). If this matching completely specified all of the fields and their first derivatives at this brane then the solution obtained by integration would completely determine the value of the fields and their radial derivatives at the second brane, and in general these need not be consistent with what would be obtained by matching to this second brane.

But matching to the first brane typically specifies only the derivatives of the fields at the first brane, and not separately the values of the fields themselves.\footnote{Since in general the bulk fields can diverge at the brane positions, this argument should more precisely be made very near to, and not precisely at, the position of the first brane.} Consequently the values of the fields at the first brane can be adjusted to try to allow the solution to properly match to the properties of the second brane. It is in this way that the system can force $\varphi_0 = \varphi_\star$ if the brane actions do not preserve the bulk shift symmetry.

{}From the perspective of a low-energy 4D observer the energy cost responsible for this stabilization looks like a scalar potential for $\varphi_0$, and our goal in what follows is to compute its shape for configurations in the immediate neighborhood of $\varphi_0 = \varphi_\star$. As the above arguments show, a classical solution subject to our assumed ansatz should not exist as soon as $\varphi_0 \ne \varphi_\star$, and the part of the ansatz responsible is likely to be the condition of maximal symmetry in the on-brane directions. No maximally symmetric solution should exist for $\varphi_0 \ne \varphi_\star$ because this indicates the onset of time evolution in response to no longer sitting at the minimum of the 4D effective potential. (This development of time dependence in response to changes in the properties of mutually gravitating brane sources resembles what happens for a system of electric charges, which generically becomes time dependent when an equilibrium arrangement is disturbed).

Rather than trying to solve for the system's time-dependent response (see however refs.~\cite{TimeDep,stability}) when $\varphi_0 \ne \varphi_\star$ we instead focus on computing features of the low-energy potential that is responsible. We do so -- in both the 4D and 6D theories -- through the artifice of turning on a current that stops the time evolution, and so removes the obstruction to static solutions. Since sufficiently small deviations from equilibrium should precipitate motion only along the low-energy flat directions, it suffices to couple the current only to these low-energy modes. By computing the amount of current required as a function of the low-energy scalar mode, we may Legendre transform in the usual way to determine the shape of the effective potential.

\subsubsection*{Linearized equations with currents}

Since we work within a linearized approximation, we perturb the brane properties in a way that does not drive the low-energy scalar fields far from their initial values. This can be achieved if the potential energy of each brane has a minimum as a function of $\phi$, and although the two branes need not agree on where this minimum is they should not disagree by too much. It suffices therefore to study the brane tensions in the vicinity of these minima, restricting to quadratic expansions in powers of $\phi$. Writing $T_b = T + \delta T_b(\phi)$, we take
\be
 \delta T_b(\phi) = T_{b0} + \frac{T_{b 2}}{2} \, (\phi - \hat \pphi_b)^2  \,,
\ee
with $b = N$ and $S$ and $T_b(\phi) = \tau_b(\phi) - \cQ \, \Phi_b(\phi)$, as before. For technical reasons --- see Appendix \ref{App:branesvsflux} --- we require that the minimum of the sum of the brane actions, $\sum_bT_b'=0$, agrees with the minimum of the sum of the fluxes, $\sum_b\Phi_b'=0$. When $\hat \pphi_\ssN \ne \hat \pphi_\ssS$ the two branes differ on which value for $\phi$ they prefer, and we assume that this difference is not so large as to invalidate a linearized integration of the field equations.

In the higher-dimensional theory the current used to stabilize the solutions against rolling is\footnote{The additional coupling $J \, \varphi_\star$ is here inserted to ensure that $J$ couples only to the light scalar mode, $\delta \varphi = \varphi_0 - \varphi_\star$, at the linearized level, and not also to the metric fluctuations. We keep this term even though, as discussed in Appendix \ref{App:misalignedJ}, for axions much lighter than the KK scale, $m \ll 1/L$, a misalignment that included metric modes only introduces subdominant contributions to the axion mass, of order $\delta m^2 \simeq m^4 L^2$.}
\be
 S_\ssJ := - \int \exd^6x \, \sqrt{-g} \; J (\phi - \varphi_\star) \,,
\ee
and, to the extent that it suffices to stabilize just the KK zero mode, $J$ can be taken to be independent of the extra-dimensional coordinates $\rho$ and $\theta$. In the presence of such a current the field equation for $\phi$ becomes
\be \label{KGwithJ}
 \partial_\ssM \Bigl( \sqrt{-g} \; g^{\ssM \ssN} \partial_\ssN \phi \Bigr)
 =  \sqrt{-g} \; \kappa^2_a J \,.
\ee

Perturbing around the rugby ball solution, our interest is in the lowest nontrivial order in $J$, corresponding to situations where the brane tensions only cause controllably small changes in $\phi$. In this case the leading approximation to the axion fluctuation is obtained by solving eq.~\pref{KGwithJ} with the metric evaluated at the rugby ball background,
\be
 \Bigl[ \sin\left( \frac{\rho}{L} \right)
 \delta \phi' \Bigr]' = \frac{\epsilon_\ssJ}{L^2} \; {\sin \left( \frac{\rho}{L}
 \right)} \,,
\ee
where the last equality defines the dimensionless current, assumed small: $\epsilon_\ssJ := \kappa^2_a J L^2 \ll 1$.

The Maxwell equation is unchanged by the current, and integrates to give
\be
 F_{\rho\theta} = A_\theta' = (\cQ + \delta\cQ) e^{-4W + B} \,.
\ee
Recall that both $W$ and $B$ in this expression include perturbations.

Including the stress energy from the current interaction, the linearized Einstein equations become
\ba
 \frac{\hat \cR}4 + \delta W'' + \frac{\delta W'}{L}
 \, \cot \left( \frac \rho L \right)
 -\frac1{2L^2} \left( \frac{\delta\cQ} \cQ \right) +
 \frac{2\delta W}{L^2} + \frac{\lambda^2 \epsilon_\ssJ
 (\phi - \varphi_\star)}{2L^2} = 0 \nn\\
 4\delta W'' + \delta B'' + \frac{2 \delta B'}{L}\, \cot \left( \frac \rho L \right)
 + \lambda^2 (\phi')^2 + \frac3{2L^2} \left( \frac{\delta\cQ} \cQ \right)
 - \frac{6\delta W}{L^2} + \frac{\lambda^2 \epsilon_\ssJ
 (\phi - \varphi_\star)}{2L^2} = 0 \nn\\
 \delta B'' + \frac{2 \delta B'}{L} \, \cot \left( \frac \rho L \right)
 + 4\delta W'  \cot\left( \frac \rho L \right)
 +\frac3{2L^2} \left( \frac{\delta\cQ}{\cQ} \right)
 -\frac{6\delta W}{L^2} +
 \frac{\lambda^2 \epsilon_\ssJ (\phi - \varphi_\star)}{2L^2} = 0 \,,
\ea
where, as before, $\lambda = \kappa/\kappa_a = \kappa F$.

\subsubsection*{Linearized solutions}

We now solve those equations to order $\epsilon_\ssJ$. Because $(\phi')^2$ is order $\epsilon_\ssJ^2$, to order $\epsilon_\ssJ$ the equation for the warping is unchanged from previous sections, giving the solution
\be
 \delta W = W_1 \cos \left( \frac\rho L \right) \,.
\ee

The current forces the axion to acquire a profile (which is desirable because this allows it to satisfy the new boundary conditions at the brane positions). The perturbed axion equation integrates to give
\be \label{phisoln}
 \delta \phi = \varphi_0 + \varphi_1 \ln \left|
 \frac{1-\cos(\rho/L)}{\sin(\rho/L)} \right|
 - \epsilon_\ssJ \ln \left| \sin \left( \frac \rho L \right) \right| \,,
\ee
with $\varphi_1$ and $\varphi_0$ integration constants. Since $\varphi_0$ parameterizes the (previously) flat direction we solve $\epsilon_\ssJ$ and all other integration constants in terms of it and brane properties.

Using these in the Einstein equations as before gives $\delta B$ as the solution to
\be
 \frac{\left[ \delta B' \sin^2 \left( \rho/ L\right)\right]'}{\sin^2
 \left( \rho/ L\right)} = \frac{10 W_1}{L^2} \, \cos \left( \frac \rho L \right)
 - \frac{3}{2L^2} \left( \frac{\delta \cQ}{\cQ} \right)
 - \frac{\lambda^2 \epsilon_\ssJ (\varphi_0 - \varphi_\star)}{2L^2}  \,,
\ee
giving
\be
 \delta B = \left[ \frac34 \left( \frac{\delta\cQ}\cQ \right)
 + \frac{\lambda^2 \epsilon_\ssJ }4 (\varphi_0 - \varphi_\star) \right] \frac\rho L \,
 \cot \left( \frac \rho L \right)
 - \frac{10W_1}3 \, \cos\left( \frac \rho L \right) + \delta B_0 \,.
\ee

Using this in the linearized flux-quantization condition finally gives a relation between $\delta \cQ$ and $\delta B_0$,
\be \label{fluxquantznlin}
 \frac{\delta\cQ}\cQ = \lambda^2 \epsilon_\ssJ (\varphi_0 - \varphi_\star) - 4 \delta B_0
 -\frac{\kappa^2 \cQ}{\pi\alpha} \left( \delta\Phi_\ssN + \delta\Phi_\ssS \right) \,.
\ee
As before, the remaining integration constants --- in this case $\varphi_1$, $W_1$ and $\delta B_0$ --- are determined by solving the matching conditions at the brane positions. The fractional change in the proper distance between the source branes becomes
\be
 \frac{\delta L}{L} = - \frac34 \left( \frac{\delta\cQ}\cQ \right)
 - \frac{\lambda^2 \epsilon_\ssJ }{4} (\varphi_0 - \varphi_\star)\,.
\ee

\subsubsection*{Matching to branes}

The matching condition for the axion at each brane is
\be
 \lim_{\rho \to \rho_b} \alpha \rho \, \phi' = \left. \frac{\kappa_a^2 T'_b(\phi)}{2\pi}
 \right|_{\rho \to \rho_b} \,,
\ee
but an additional complication arises because the right-hand side is ill defined due to the  divergence in $\phi(\rho)$ at the brane positions. This requires a renormalization of the parameters defining the brane potentials \cite{otheruvcaps, branerenormalization}. To this end first regularize the matching condition by evaluating it at $\rho = \rho_\ssN + \varepsilon_\ssN$ and $\rho = \rho_\ssS - \varepsilon_\ssS$. Then define the renormalized parameters
\ba
 \pphi_{\ssN} := \hat \pphi_\ssN - (\varphi_1 - \epsilon_\ssJ )
 \ln\left( \frac{\varepsilon_\ssN} L \right) + \varphi_1 \ln 2 \nn\\
 \pphi_{\ssS} := \hat \pphi_\ssS + (\varphi_1 + \epsilon_\ssJ )
 \ln\left( \frac{\varepsilon_\ssS} L \right) - \varphi_1 \ln 2 \,.
\ea
Because the field profile satisfies
\be
 \phi(\varepsilon_\ssN) = \varphi_0 + \varphi_1 \ln \left(
 \frac{\varepsilon_\ssN}{2L} \right)
 - \epsilon_\ssJ \ln \left( \frac {\varepsilon_\ssN} L \right) \,,
\ee
(and a similar result at $\rho = \pi L - \varepsilon_\ssS$), these definitions ensure
\be
 \phi(\varepsilon_b) - \hat \pphi_b = \varphi_0 - \pphi_{b} \,,
\ee
and so remain finite in the limit $\varepsilon_b \to 0$. This makes the derivative of the tension (and the tension itself) finite when evaluated on the brane. Because the fluxes are also written in terms of $\phi-\hat \pphi_b$, they do not need a separate renormalization.

In terms of renormalized quantities the matching conditions directly relate the integration constants,
\ba
 \alpha( \varphi_1 - \epsilon_\ssJ) &=& \frac{ \kappa_a^2 \delta T_\ssN'(\varphi_0)}{
 2 \pi} = \left( \frac{\kappa_a^2  T_{\ssN 2}}{2\pi}
 \right) \left( \varphi_0 - \pphi_\ssN \right) \nn\\
 -\alpha(\varphi_1 + \epsilon_\ssJ) &=&  \frac{ \kappa_a^2 \delta T_\ssS'(\varphi_0)}{
 2 \pi} = \left( \frac{\kappa_a^2 T_{\ssS 2}}{2\pi} \,
 \right) \left(\varphi_0 - \pphi_{\ssS}\right) \,,
\ea
allowing the inference
\ba
\label{current relation}
 \epsilon_\ssJ &=& - \frac{\kappa_a^2}{4\pi\alpha} \Bigl[
 \delta T_\ssN'(\varphi_0) + \delta T_\ssS'(\varphi_0) \Bigr]
 = - \frac{\kappa_a^2}{4\pi\alpha} \Bigl[ ( T_{\ssN 2}
 + T_{\ssS 2} ) \, \varphi_0 - T_{\ssN 2} \, \pphi_\ssN
 - T_{\ssS 2} \, \pphi_\ssS \Bigr] \nn\\
 \varphi_1 &=& \frac{\kappa_a^2}{4\pi\alpha} \Bigl[
 \delta T_\ssN'(\varphi_0) - \delta T_\ssS'(\varphi_0) \Bigr]
 = \frac{\kappa_a^2}{4\pi\alpha} \Bigl[ (T_{\ssN 2}
 - T_{\ssS 2} ) \, \varphi_0 + T_{\ssS 2} \, \pphi_\ssS
 - T_{\ssN 2} \, \pphi_\ssN \Bigr] \,.
\ea
The first of these identifies the field value where the flat direction gets stabilized, $\varphi_0 = \varphi_\star$, since this is the solution that corresponds to zero external current. The condition $\epsilon_\ssJ(\varphi_\star) = 0$ implies $\varphi_\star$ satisfies
\be \label{6Dstnrypt}
 \delta T_\ssN'(\varphi_\star) + \delta T_\ssS' (\varphi_\star) = 0 \,,
\ee
and so when $\delta T_b(\varphi_0) = T_{b 0} + \frac12 \, T_{b 2} \, (\varphi_0 - \pphi_b)^2$
\be
 \varphi_\star = \frac{T_{\ssN 2} \pphi_\ssN + T_{\ssS 2} \pphi_\ssS}{T_{\ssN 2} +
 T_{\ssS 2}} \,.
\ee

We again fix $\delta B_0$ and $W_1$ from the last of the matching conditions, eqs.~\pref{matching},
\be
 \left(e^B\right)'_{\rho_b} = 1 - \frac{\kappa^2}{2\pi} \Bigl[
 T +  \delta T_b (\varphi_0 ) \Bigr] \,,
\ee
which uses $\cU_b (\varphi_0) \simeq 0$, as can be inferred either from the second of eqs.~\pref{matching}, or by solving eq.~\pref{curvature constraint} to linear order in $\kappa^2 T_b$. As before this leads to the conditions
\ba
 \alpha \left[ -2\delta B_0 - \frac{10 W_1}3 - \frac{3\kappa^2 \cQ
 \delta\Phi_{\rm tot}}{4\pi\alpha} + \lambda^2 \epsilon_\ssJ (\varphi_0 - \varphi_\star) \right]
 &=& - \left( \frac{\kappa^2}{2\pi} \right) \delta T_\ssN(\varphi_0)  \nn\\
 \alpha \left[ -2 \delta B_0 + \frac{10 W_1}3 - \frac{3\kappa^2 \cQ
 \delta\Phi_{\rm tot}}{4\pi\alpha} + \lambda^2 \epsilon_\ssJ (\varphi_0 - \varphi_\star) \right]
 &=& - \left( \frac{\kappa^2}{2\pi} \right) \delta T_\ssS ( \varphi_0 ) \,.
\ea
The result is
\be
 W_1 = \frac{3\kappa^2}{40 \pi \alpha}
 \Bigl[ \delta T_\ssN ( \varphi_0) -\delta T_\ssS (\varphi_0) \Bigr]  \,,
\ee
and
\ba
 \delta B_0 &=& \frac{\lambda^2 \epsilon_\ssJ}2 (\varphi_0 - \varphi_\star)
 + \frac{\kappa^2}{8\pi\alpha} \Bigl\{ \delta T_\ssN (\varphi_0)
 +\delta T_\ssS (\varphi_0) - 3 \cQ \Bigl[ \delta\Phi_\ssN (\varphi_0)
 + \delta\Phi_\ssS(\varphi_0) \Bigr] \Bigr\} \\
  &=& \frac{\kappa^2}{8\pi\alpha} \left\{ \sum_{b = \ssN,\ssS}
  \Bigl[ \delta T_b (\varphi_0)-
  (\varphi_0 - \varphi_\star) \delta T_b'(\varphi_0) \Bigr]
  - 3 \cQ  \delta\Phi_{\rm tot}
   (\varphi_0) \right\} \,, \nn
\ea
where the final line eliminates $\epsilon_\ssJ$ using eq.~\pref{current relation} and $\lambda^2 \kappa_a^2 = \kappa^2$. This implies
\ba
 \frac{\delta \cQ}{\cQ} &=& - \lambda^2 \epsilon_\ssJ (\varphi_0 - \varphi_\star)
 - \frac{\kappa^2}{2\pi\alpha} \Bigl\{ \delta T_\ssN (\varphi_0)
 +\delta T_\ssS (\varphi_0) - \cQ  \delta\Phi_{\rm tot} (\varphi_0)  \Bigr\} \\
 &=&  - \frac{\kappa^2}{2\pi\alpha} \left\{ \sum_{b = \ssN, \ssS} \left[
 \delta T_b (\varphi_0) - \frac12 (\varphi_0 - \varphi_\star) \, \delta T_b'(\varphi_0)
 \right] - \cQ \delta\Phi_{\rm tot} (\varphi_0) \right\}  \,. \nn
\ea

In terms of these the fractional change in the proper distance between branes becomes
\ba
 \frac{\delta L}{L} &=& - \frac34 \left( \frac{\delta\cQ}\cQ \right)
 - \frac{\lambda^2 \epsilon_\ssJ}{4} (\varphi_0 - \varphi_\star) \nn\\
 &=& \frac{ \lambda^2 \epsilon_\ssJ}{2} (\varphi_0 - \varphi_\star)
 + \frac{3\kappa^2}{8\pi\alpha} \Bigl\{ \delta T_\ssN (\varphi_0)
 +\delta T_\ssS (\varphi_0) - \cQ  \delta\Phi_{\rm tot} (\varphi_0)  \Bigr\} \nn\\
 &=& \frac{3\kappa^2}{8\pi\alpha} \left\{ \sum_{b = \ssN, \ssS} \left[
 \delta T_b (\varphi_0) - \frac13 (\varphi_0 - \varphi_\star) \, \delta T_b'(\varphi_0)
 \right] - \cQ \delta\Phi_{\rm tot} (\varphi_0) \right\}\,.
\ea

\subsubsection*{On-brane geometry and 4D effective potential}

The linearized Einstein equation yields the following on-brane curvature
\ba \label{Rhatvsphi0}
 \hat \cR &=& -4 \left[ \delta W'' + \frac{\delta W'}{L} \,
 \cot \left( \frac \rho L \right) \right]
 - \frac{8 \delta W }{L^2} + \frac{2}{L^2} \left( \frac{\delta\cQ}{\cQ} \right)
 - \frac{2 \lambda^2 \epsilon_\ssJ }{L^2} (\varphi_0 - \varphi_\star) \nn\\
 &=& \frac{2}{L^2} \left[ \frac{\delta\cQ}\cQ  - \lambda^2
 \epsilon_\ssJ (\varphi_0 - \varphi_\star) \right] \\
  &=& -\frac{\kappa^2}{\pi\alpha L^2} \left\{ \sum_{b = \ssN, \ssS}
  \Bigl[ \delta T_b (\varphi_0) -
  (\varphi_0 - \varphi_\star) \, \delta T_b'(\varphi_0) \Bigr]
  - \cQ  \delta\Phi_{\rm tot}
   (\varphi_0) \right\} \,. \nn
\ea

The presence of the current $J$ complicates the determination of the effective potential, $V_{\rm eff}(\varphi)$, in the low-energy 4D theory. The appropriate matching calculation turns on a current in the low-energy theory as well, and asks what potential reproduces the previous results for $\hat \cR$ and $\varphi_\star$.

The most general action for the 4D effective theory involving only the 4D metric, $\hat g_{\mu\nu}$, and the low-energy scalar, $\varphi$, is (up to two derivatives)
\be
 S_{\rm eff} = - \int \exd^4x \, \sqrt{- \hat g} \; \left[
 \frac1{2 \kappa_4^2} \,  \hat g^{\mu\nu} \Bigl( \hat \cR_{\mu\nu}
 + \lambda^2 \, \pd_\mu \varphi \, \pd_\nu \varphi
 \Bigr) + V_{\rm eff}(\varphi) + j (\varphi - \varphi_\star) \right] \,,
\ee
where $j$ is the low-energy current, $\kappa_4$ is given by eq.~\pref{4DkappaDR} and the 4D axion decay constant is
\be
 f^2 = 4 \pi \alpha L^2 F^4 = \frac{4 \pi \alpha L^2}{\kappa_a^2} = \frac{\lambda^2}{\kappa^2_4} \,.
\ee
We couple the current $j$ to the difference $\varphi - \varphi_\star$ purely as a matter of later convenience.

The equations of motion, specialized to constant scalar fields, $\varphi = \varphi_0$, and to maximally symmetric geometries, are
\be
 j = - V_{\rm eff}'(\varphi_0)
 \quad \hbox{and} \quad
 \frac{\hat \cR}{4 \kappa_4^2} = -j (\varphi_0 - \varphi_\star) - V_{\rm eff}(\varphi_0) \,,
\ee
from which $j$ can be eliminated to give
\be \label{4DRvsVeff}
 (\varphi_0 - \varphi_\star) V_{\rm eff}'(\varphi_0) - V_{\rm eff}(\varphi_0)
 = \frac{\hat \cR}{4 \kappa_4^2}  = \frac{\pi \alpha L^2 \hat \cR}{\kappa^2} \,.
\ee

The functional form for the potential $V_{\rm eff}$ is determined by requiring eq.~\pref{4DRvsVeff} to reproduce the curvature, eq.~\pref{Rhatvsphi0}, predicted by the 6D theory, regarded as a function of $\varphi_0$. This can be obtained by regarding eq.~\pref{4DRvsVeff} as a differential equation for $V_{\rm eff}$, whose solution is
\ba \label{Veffsolngen}
 V_{\rm eff}(\varphi_0) &=& (\varphi_0 - \varphi_\star)
 \int \frac{\exd \varphi}{(\varphi - \varphi_\star)^2} \;
 \left[ \frac{ \pi \alpha L^2 \hat \cR(\varphi)}{\kappa^2} \right] \\
 &=& \delta T_\ssN(\varphi_0) + \delta T_\ssS(\varphi_0) +
  (\varphi_0 - \varphi_\star) \left\{ \int_{\varphi_\star}^{\varphi_0} \exd \varphi
 \; \left[ \frac{\cQ  \delta\Phi_{\rm tot}(\varphi)}{(\varphi - \varphi_\star)^2}  \right]
 - \lim_{\varphi \to \varphi_\star} \left[ \frac{\cQ \delta\Phi_{\rm tot}(\varphi)}{
 \varphi - \varphi_\star} \right] \right\} \,.\nn
\ea
In general, the coefficient of the term linear in $(\varphi_0 - \varphi_\star)$ in $V_{\rm eff}$ is the integration constant, which is fixed in the second equality of eq.~\pref{Veffsolngen} by requiring $V_{\rm eff}'(\varphi_\star) = 0$, for $\varphi_\star$ as given in the 6D theory by eq.~\pref{6Dstnrypt}.

Two physical parameters of particular interest here are: ($i$) the effective on-brane cosmological constant, $\vaceng := V_{\rm eff}(\varphi_\star)$, and the low-energy scalar mass, $m_\varphi^2 := V''_{\rm eff}(\varphi_\star)/f^2$. The first of these evaluates to
\ba
 \vaceng := V_{\eff}(\varphi_\star)
  &=& \delta T_\ssN(\varphi_\star) + \delta T_\ssS (\varphi_\star)
  -  \cQ \Bigl[ \delta\Phi_\ssN(\varphi_\star)+ \delta\Phi_\ssS(\varphi_\star)
  \Bigr] \nn\\
  &=& \delta \tau_\ssN(\varphi_\star) + \delta \tau_\ssS (\varphi_\star)
  - 2 \cQ \Bigl[ \delta\Phi_\ssN(\varphi_\star)+ \delta\Phi_\ssS(\varphi_\star)
  \Bigr]\,,
\ea
whose value agrees with our earlier Einstein-Maxwell calculation when $\delta T_b' = 0$. Similarly\footnote{Notice that $m^2_\varphi$ diverges if $\Phi_{\rm tot}$ vanishes linearly with $\varphi-\varphi_\star$. In this case the lowest energy KK mode is not properly captured by our ansatz --- see Appendix \ref{App:branesvsflux} --- and so the low-energy potential misidentifies its size. It is for this reason that we require $\Phi_{\rm tot}$ and $T_{\rm tot}$ to agree on the value $\varphi_\star$ at which they are minimized.}
\be
 m_\varphi^2 = \frac{V''_{\rm eff}(\varphi_\star)}{f^2}
 = \frac{1}{f^2} \left[ \delta T_\ssN''(\varphi_\star)
 + \delta T_\ssS''(\varphi_\star) + \lim_{\varphi \to \varphi_\star}
 \left(  \frac{ \cQ \delta\Phi_\ssN'(\varphi) + \cQ
 \delta\Phi_\ssS'(\varphi)}{\varphi - \varphi_\star}
 \right) \right] \,.
\ee
In the absence of brane fluxes the effective potential is simply the sum of the brane potentials. But although the low-energy scalar always stabilizes at the stationary points of $\sum_b \delta T_b$, the scalar masses and 4D cosmological constant in general differ from what would be expected based just on $\sum_b \delta T_b$.

\subsubsection*{Comparison with dimensional reduction}

As before, we can also evaluate $V_{\rm eff}$ at the classical level by direct dimensional reduction, which gives the integral
\ba
 V_{\rm eff} &=& 2\pi \int_0^{\pi L} \exd \rho \; e^{B+4W} \left\{ \frac1{2\kappa^2}
 \left[ \frac{8W'}{L} \, \cot \left( \frac\rho L \right) \right] -\frac14 (\cQ + \delta\cQ)^2 e^{-8W} + \frac12 \left( \Lambda + J\phi\right) \right\} \nn\\
 &=& \frac{\pi\alpha L^2}{\kappa^2} \int_0^{\pi L} \frac{\exd\rho}{L} \,
 \sin \left( \frac \rho L \right)
 \left[ -\kappa^2 \cQ^2 \left( \frac{\delta\cQ}\cQ \right)
 + \kappa^2 J \phi \right] \nn\\
 &=& -\frac{2 \pi\alpha}{\kappa^2} \left( \frac{\delta\cQ}\cQ - \lambda^2\epsilon_\ssJ \varphi_0 \right)
 = - \frac{\pi \alpha L^2}{\kappa^2} \, \hat \cR \,,
\ea
in agreement with the 6D calculation above.

\section{Applications and special cases}

This section seeks to illustrate the physical implications of the previous section's results by exploring several instructive examples.

\subsection{Bulk response to stabilizing potentials}

Consider first the response of the bulk geometry and the properties of the low-energy 4D scalar-tensor theory, distinguishing the cases where the two brane agree on, or compete for, the field value where the low-energy scalar is stabilized.

\subsubsection*{Shared minima}

As an example where the fluxes and tensions on both branes are minimized at a common value of $\varphi_0$, consider the special case that all the fluxes and tensions have the following expansion $\tau_b = T + \delta \tau_b(\phi)$ and $\Phi_b = \Phi + \delta \Phi_b(\phi)$ with
\be
 \delta \tau_b(\varphi_0) = \delta \tau_{b 0}
 + \frac{\delta \tau_{b 2}}{2} ( \varphi_0 - \pphi)^2
\quad \hbox{and} \quad
 \delta \Phi_b(\varphi_0) = \delta \Phi_{b0}
 + \frac{\delta \Phi_{b2}}{2} ( \varphi_0 - \pphi)^2 \,,
\ee
and so
\be
 \delta T_b(\varphi_0) = \delta \tau_b(\varphi_0) - \cQ \delta \Phi_b(\varphi_0)
 = \delta T_{b 0} + \frac{\delta T_{b 2}}{2} ( \varphi_0 - \pphi)^2 \,,
\ee
where $\delta T_{b k} = \delta \tau_{b k} - \cQ \, \delta \Phi_{b k}$ are constants.

The condition fixing $\varphi_\star$ in this case is $\varphi_\star = \pphi$, as one would expect. Inserting this into the formulae for the relative warping of the two branes and the fractional change in inter-brane distance gives
\ba
 \delta W_\ssN - \delta W_\ssS &=& \frac{3\kappa^2}{20 \pi \alpha}
 \Bigl[ \delta  T_{\ssN 0}  - \delta T_{\ssS 0} \Bigr] \nn\\
 \frac{\delta L}{L} &=& \frac{3\kappa^2}{8\pi\alpha} \left[ \Bigl(
 \delta T_{\ssN 0} + \delta T_{\ssS 0} \Bigr) - \cQ  \Bigl( \delta \Phi_{\ssN 0} +
 \delta \Phi_{\ssS 0} \Bigr) \right] \,.
\ea
Similarly, the on-brane expressions for $\vaceng$ and $m^2_\varphi$ yield
\ba \label{Lambdamexamp1}
 \vaceng &=& \delta T_{\ssN 0} + \delta \delta T_{\ssS 0} - \cQ \Bigl(
 \delta \Phi_{\ssN 0} + \delta \Phi_{\ssS 0} \Bigr) \nn\\
 &=&  \delta \tau_{\ssN 0} + \delta \tau_{\ssS 0} - 2\cQ \Bigl(
 \delta \Phi_{\ssN 0} + \delta \Phi_{\ssS 0} \Bigr)  \,,
\ea
and
\ba \label{Lambdamexamp2}
 m_\varphi^2 &=&  \frac{ 1}{f^2} \Bigl[\delta  T_{\ssN 2} + \delta T_{\ssS 2}
 + \cQ \Bigl( \delta \Phi_{\ssN 2} + \delta \Phi_{\ssS 2} \Bigr) \Bigr]  \nn\\
 &=&  \frac{ 1}{f^2} \Bigl[ \delta \tau_{\ssN 2} + \delta \tau_{\ssS 2} \Bigr]  \,.
\ea
Notice that only the second derivative of the tension, $\delta \tau_b''(\varphi_\star)$, contributes to the scalar mass, while both the tension, $\delta \tau_b(\varphi_\star)$, and the flux, $\delta \Phi_b(\varphi_\star)$, contribute to the on- and off-brane curvatures.

\subsubsection*{Brane competition}

Consider next the case where the two branes each prefer $\varphi_0$ to stabilize at different values, causing them to compete in the value they ultimately determine. A representative example in this case is
\ba
 \delta T_b(\varphi_0) &=& \delta  T_{b 0} + \frac{\delta T_{b 2}}{2}
 \, ( \varphi_0 - \pphi_b)^2 \nn\\
 \delta \Phi_b(\varphi_0) &=& \delta \Phi_{b0} +
  \frac{\delta \Phi_{b2}}{2} \, ( \varphi_0
 - \pphi_b)^2 \,.
\ea
The stabilizing value for the scalar is now neither $\pphi_\ssN$ nor $\pphi_\ssS$, but instead the intermediate value
\be
 \varphi_\star = \frac{\delta T_{\ssN 2} \pphi_\ssN
 + \delta T_{\ssS 2} \pphi_\ssS}{\delta T_{\ssN 2} + \delta T_{\ssS 2}} \,,
\ee
with the ratio $\delta T_{\ssN 2}/\delta T_{\ssS 2}$ controlling precisely where $\varphi_\star$ lies between $\pphi_\ssN$ and $\pphi_\ssS$. Requiring $\delta \Phi_{\rm tot} = \delta \Phi_\ssN + \delta \Phi_\ssS$ also to have its minimum at the same value of $\varphi_\star$ then requires
\be \label{equalratios}
 \frac{\delta \Phi_{\ssN 2}}{\delta \Phi_{\ssS 2}}
 = \frac{ \delta T_{\ssN 2} }{\delta  T_{\ssS 2}} \,.
\ee

The extra-dimensional geometry satisfies
\be
 \delta W_\ssN - \delta W_\ssS =
 \frac{3 \kappa^2}{20 \pi \alpha} \left\{ \delta  T_{\ssN 0} -
 \delta T_{\ssS 0} + \frac{\delta T_{\ssN 2} \delta T_{\ssS 2}}2
 \left[ \frac{\delta T_{\ssS 2} - \delta T_{\ssN 2}}{(\delta T_{\ssN 2}
 + \delta T_{\ssS 2})^2} \right] (\pphi_\ssN - \pphi_\ssS)^2  \right\} \,,
\ee
and $\delta L/L = (3 \kappa^2 \vaceng/8\pi \alpha)$, with the 4D vacuum energy given by
\ba
 \vaceng &=& \delta T_{\ssN 0} + \delta T_{\ssS 0}
 - \cQ \Bigl(\delta  \Phi_{\ssN 0} + \delta \Phi_{\ssS 0} \Bigr) + \frac12
 \left( \frac{ \delta T_{\ssN 2} \delta T_{\ssS 2}}{ \delta T_{\ssN 2}
 +\delta  T_{\ssS 2}} \right)
 (\pphi_\ssN - \pphi_\ssS )^2 \nn\\
 && \qquad \qquad \qquad
  - \frac{\cQ}{2} \left[ \frac{\delta \Phi_{\ssN 2} \delta T_{\ssS 2}^2
 + \delta \Phi_{\ssS2} \delta T_{\ssN 2}^2}{ (
 \delta T_{\ssN 2} + \delta T_{\ssS 2})^2} \right]
 (\pphi_\ssN-\pphi_\ssS)^2  \nn\\
 &=& \delta T_{\ssN 0} +  \delta T_{\ssS 0}
 - \cQ \Bigl(\delta  \Phi_{\ssN 0} + \delta \Phi_{\ssS 0} \Bigr) + \frac12
 \left[ \frac{ \delta T_{\ssN 2} (\delta T_{\ssS 2}
 - \cQ \delta \Phi_{\ssS 2})}{ \delta T_{\ssN 2} +\delta  T_{\ssS 2}} \right]
 (\pphi_\ssN - \pphi_\ssS )^2 \,,
\ea
where the last equality uses eq.~\pref{equalratios}. The result for $m^2_\varphi$ is again given by eq.~\pref{Lambdamexamp2}. Because $\varphi_\star$ does not minimize the tension at either brane both the total tension and total flux get increased by positive amounts. These positive contributions then act oppositely in $\vaceng$.

More complicated competitions can also occur if there is also symmetry-breaking in the bulk, in which case competition between the bulk and brane potentials can lead to self-localization \cite{selfloc}.

\subsubsection*{Flux domination}

A particular instance of the previous scenario corresponds to the case where $| \delta \tau_b| \ll |\cQ \delta\Phi_b|$, since in this case $\delta T_b(\varphi) \simeq - \cQ \delta\Phi_b(\varphi)$. Then the stabilizing value for the scalar becomes
\be
 \varphi_\star = \frac{\delta \Phi_{\ssN 2} \pphi_\ssN
 + \delta \Phi_{\ssS 2} \pphi_\ssS}{\delta \Phi_{\ssN 2} +
 \delta \Phi_{\ssS 2}} \,,
\ee
and
\be
 \delta W_\ssN - \delta W_\ssS = -\frac{3 \kappa^2 \cQ}{20 \pi \alpha}
 \left\{ \delta \Phi_{\ssN 0} - \delta \Phi_{\ssS 0}
 + \frac{\delta \Phi_{\ssN 2} \delta \Phi_{\ssS 2}}2 \left[
 \frac{\delta \Phi_{\ssS 2} - \delta \Phi_{\ssN 2}}{(\delta \Phi_{\ssN 2}
 + \delta \Phi_{\ssS 2})^2} \right]
 (\pphi_\ssN - \pphi_\ssS)^2  \right\} \,,
\ee
while $\delta L/L = (3 \kappa^2 \vaceng/8\pi \alpha)$, with
\be
 \vaceng =  - 2\cQ \Bigl(\delta  \Phi_{\ssN 0} + \delta \Phi_{\ssS 0} \Bigr)
 - \frac{\cQ}{2} \left( \frac{ \delta \Phi_{\ssN 2} \delta \Phi_{\ssS 2}}{
 \delta \Phi_{\ssN 2}
 + \delta \Phi_{\ssS 2}} \right) (\pphi_\ssN - \pphi_\ssS )^2 \,.
\ee
Because in this case $m^2_\varphi \simeq 0$ to leading order, the scalar mass --- and so also the stability of the vacuum $\varphi_0 = \varphi_\star$ --- is controlled by subdominant effects (like $\delta \tau_b$ or loops), even though the flux dominates the classical contribution to $\vaceng$.

\subsection{Axions}

It is instructive to consider the relative sizes of the various scales that arise naturally when bulk axions receive masses through their couplings to branes, since these need not be related in the same way as when both axion and symmetry-breaking physics share the same number of dimensions. This section briefly examines several illustrative choices.

There are five scales that naturally arise in bulk-axion models. Three of these --- the extra-dimensional Planck scale, $M_g = \kappa^{-1/2}$; the axion decay constant, $F = \kappa_a^{-1/2}$; and the KK scale, $m_\KK = 1/L$ --- characterize the bulk physics. The source branes are responsible for the other two: the scale $\Lambda$ set by the $\phi$-independent parts of the brane tensions and fluxes; and the scale $\mu$ set by the $\phi$-dependent terms,
\be
 \tau_{b0} \simeq \Lambda^4 \,, \quad
 \Phi_{b0} \simeq \Lambda \,, \quad
 \tau_{b2} \simeq \mu^4
 \quad \hbox{and} \quad
 \Phi_{b2} \simeq \mu \,.
\ee

These scales are not completely arbitrary. In general, control over the semiclassical approximation requires $M_g$ to be much bigger than all of the others. Although the conditions $\kappa \Lambda^2 \ll 1$, $\kappa \mu^2 \ll 1$ and $\kappa /L^2 \ll 1$ follow fairly directly from standard arguments \cite{GREFT}, the condition $F \ll M_g$ is a bit more indirect. Because $F$ sets the scale of the bulk symmetry breaking for which $\phi$ is the would-be Goldstone boson, our upper bound on $F$ assumes the UV completion describing this breaking intercedes below the Planck scale (before which the UV completion associated with gravity --- such as string theory --- should also intercede).

Furthermore, we generically expect $\mu \lsim \Lambda$ for generic types of brane physics. This follows because it is difficult to have physics contribute to the $\varphi$ mass without also contributing equivalently to the vacuum energy. Notice in this regard that it is technically natural to take $\mu \ll \Lambda$, because it is only $\mu$ that breaks the shift symmetry of the low-energy scalar: $\varphi \to \varphi + \hbox{(constant)}$.

\subsubsection*{Axion mass}

In terms of these scales the mass of the light scalar in the effective 4D theory is of order
\be
 m_\varphi \simeq \frac{\mu^2}{f} \simeq \left( \frac{\mu}{F} \right)^2 \frac 1L
 \simeq \left( \frac{\mu}{F} \right)^2 m_\KK \,,
\ee
in all three of the scenarios considered above.\footnote{In some circumstances additional suppression can be achieved, such as if the Goldstone symmetry is not completely broken by either brane separately \cite{BGH}.} This result doesn't depend on which scenario is considered because for all three the scalar mass depends only on $\sqrt{\delta \tau_{\ssN 2} + \delta \tau_{\ssS 2}}/f$. Provided $\mu \ll F$, $\varphi$ is much lighter than the KK scale as is appropriate for its description in the low-energy 4D effective theory.

For the higher-dimensional models of interest here, however, the regime $\mu \gg F$ can also make sense. Extra dimensions allow this regime even though the scale $\mu$ of explicit symmetry breaking is then much larger than the  scale of the spontaneous breaking: $F$. Because all symmetry breaking is localized on the branes, even though $\mu > F$ the field $\phi$ behaves like a Goldstone boson for all energies lower than $F$ in the bulk provided one stays away from the position of the branes. Although this regime is not amenable to a 4D description, the mass of all KK modes {\em can} be computed within the higher-dimensional theory. In this limit the `zero mode' becomes lost among the generic massive KK states and is not singled out as being particularly light. In this regime it is clear that otherwise standard arguments, like cosmological bounds on axion properties, cannot be made purely within four dimensions without taking the full dynamics of the extra dimensions into account.

\subsubsection*{Curvatures}

A second robust prediction of all of the above scenarios is the relation between the change to the extra-dimensional size and the four-dimensional curvature:
\be
 \frac{\delta L}{L} = \frac{3\kappa^2 \vaceng}{8 \pi \alpha} \,,
\ee
although the size of $\vaceng$ itself is not as model independent. This source of this model dependence is the competition between tension and flux contributions to $\vaceng$, whose competing contributions are of order $\delta \vaceng \simeq \sum_b \tau_{b0}$ or $\delta \vaceng \simeq \sum_b \cQ \Phi_{b0}$, with
\be
 \tau_{b0} \simeq \Lambda^4
 \quad \hbox{and} \quad
 \cQ \Phi_{b0} \simeq \cQ \Lambda \simeq \frac{\Lambda}{\kappa L}
 \simeq \frac{M_g^2 \Lambda}{L} \simeq \sqrt{4\pi \alpha}
 \; \left( \frac{M_g^4 \Lambda}{M_p} \right) \,.
\ee
Special things happen for the BPS-like situation when the tension and charge are precisely related, $\tau_b(\varphi_\star) = 2 \cQ \Phi_b(\varphi_\star)$, since in this case the two contributions to $\vaceng$ precisely cancel.

Whether the tension or the flux dominates in $\vaceng$ depends on where $\Lambda$ sits relative to the two geometrical scales $M_p \simeq 10^{18}$ GeV and $1/L$. Defining $\Lambda_\star^3 := \sqrt{4\pi \alpha} \, M_g^4/M_p$ we have $\vaceng \simeq \Lambda^4$ if $\Lambda > \Lambda_\star$ and $\vaceng \simeq \Lambda  \Lambda_\star^3$ when $\Lambda < \Lambda_\star$. Some representative numerical values are given in Table 1. Intriguingly, $\Lambda_\star$ is of order the QCD scale in the extreme case of large extra dimensions ($M_g \lsim 10$ TeV and $m_\KK \lsim 0.4$ eV \cite{ADD}).

\vspace{3mm}
\begin{center}
\begin{tabular}{ccc}
              \hline
              \hline
              $M_g$ & $\Lambda_\star$ & $m_\KK$ \\
              \hline
              $10^{15}$ & $4 \times 10^{14}$ & $4 \times 10^{12}$ \\
              $10^{11}$ & $2 \times 10^{9}$ & $4 \times 10^{3}$ \\
              $10^7$ & $8 \times 10^3$ & $4 \times 10^{-4}$ \\
              $10^4$ & $0.8$ & $4 \times 10^{-10}$ \\
              \hline
              \hline
\end{tabular} \\
\vspace{1mm}
{\bf Table 1:} Values of $m_\KK$ and $\Lambda_\star$ as a function of $M_g$ (in GeV).
\end{center}

\subsection{Gravitationally coupled scalars}

The special case $F \simeq M_g$ is of particular interest because then $f \simeq M_p$ and the low-energy 4D scalar is gravitationally coupled. In this case the light scalar mass is robustly of order $m_\varphi \simeq \mu^2/M_p$, and its small size is technically natural since it is protected by the underlying shift symmetry. There are two situations for which the existence of such light weakly-coupled scalars are of particular interest.

\subsubsection*{An inflationary mechanism}

Inflationary models famously require light, weakly coupled scalars; something that is usually fairly difficult to achieve without fine-tuning in a real microscopic theory. The above estimates point to a fairly generic mechanism for achieving slow-roll inflation whenever a bulk axion acquires a potential through its interaction with codimension-two branes. This mechanism can be regarded as an ultraviolet completion of 4D `natural inflation' models \cite{NatInf}, that assume the inflaton to be a pseudo-Goldstone particle.

The mechanism rests on two assumptions: ($i$) the brane energy density, $\vaceng$, must dominate any other contributions to the geometry in the on-brane directions; and ($ii$) the brane-axion couplings must have a local {\em maximum} rather than a minimum at $\varphi = \varphi_\star$, for which $m_\varphi^2$ is of order $\mu^2/M_p$ (as above) but negative. In this case because the previous estimates apply near the potential's maximum, with an effective 4D scalar potential being of order
\be
 V_\eff(\varphi) \simeq A + B U \left( \varphi - \varphi_\star \right) \,,
\ee
with $B \simeq \cO(\mu^4)$ and $A = \vaceng \simeq \cO(\Lambda^4)$ or $\vaceng \simeq \cO( \Lambda M_g^4/M_p)$ (whichever is larger). The technically natural choice $\mu \ll \Lambda$, $(M_g^4/M_p)^{1/3}$ therefore ensures $B \ll A$. Here $U(x)$ is a calculable, dimensionless, order-unity function, whose expansion for small arguments is (by assumption) $U(x) \simeq - \frac 12 \, U_2 \, x^2 + \cdots$ with $U_2 > 0$ and order unity.

Should this potential dominate the 4D geometry it produces a Hubble scale near this maximum that is of order $H \simeq \sqrt\vaceng/M_p$ and so $H$ is of order the larger of $\Lambda^2/M_p$ or $(M_g^2/M_p) (\Lambda/M_p)^{1/2}$. Because of this, our choice $\mu \ll \Lambda$ automatically ensures $|m_\varphi^2| \ll H$. Provided that $H$ is also small compared with the KK scale --- as is easy to arrange --- the resulting cosmology can be understood within the 4D effective theory, and describes an inflationary slow roll provided $\varphi$ starts in an initially spatially homogeneous configuration near the potential's maximum. This slow roll is inflationary (despite having $f \simeq M_p$) because $B \ll A$, since the slow-roll parameters are of order $\epsilon \simeq (B U'/A)^2$ and $\eta \simeq B U''/A$. $\eta$ is sufficiently small to inflate for $\sim 60$ $e$-foldings if $B/A \simeq 0.01$, in which case $\epsilon \simeq \eta^2$ is even smaller (and so the inflation typically does not produce an observable signal of primordial gravity-waves). If $A \simeq \Lambda^4$ then $B/A \simeq (\mu/\Lambda)^4$ and a sufficiently small ratio can be ensured for the comparably modest hierarchy $\mu/\Lambda \simeq 0.3$.

As an existence proof that all parameters can be chosen as required above consider the intriguing, but extreme, scenario where the QCD axion is a bulk scalar within large extra dimensions (the last line of Table 1). In this case taking $\mu \simeq \Lambda/3 \simeq \Lambda_{\QCD} \simeq 0.2$ GeV both provides the right scale of axion-matter couplings, and ensures $\Lambda \simeq \Lambda_\star$ and so $\vaceng \simeq \Lambda^4$ and $H \ll m_\KK$. One might imagine that whatever solves the cosmological constant problem arranges the true ground state of the present epoch to be the unperturbed rugby-ball solution having $\hat \cR \simeq 0$ and $T$ of order the weak scale, with the perturbation $\delta \tau_b \simeq \Lambda_\QCD^4$ arising in the early universe due to the vacuum energy associated with the QCD phase transition on the brane. Even if it were not to involve enough $e$-foldings to account for primordial fluctuations, such a very late inflationary period could be useful for removing unwanted relics --- like moduli or KK modes --- from the much earlier universe.

\subsubsection*{New long-range forces}

Another potential application (or constraint) on the light bulk Goldstone mode described here comes from the long-range forces that it would mediate if its mass is sufficiently light. Indeed, one motivation to study the brane-bulk dynamics explored above is to find sensible UV completions which can have a technically light scalar whose presence could be sought when testing general relativity. Such tests provide strong constraints on the existence of any new forces competing with gravity in the solar system, with a precision that varies with the mass of the new scalar particle and the nature of its couplings to matter \cite{GRTests}.

This section explores what brane-bulk dynamics might say about the couplings of the low-energy scalar to matter localized on the branes. We find these couplings can (but need not, depending on the brane properties) realize some earlier-proposed mechanisms \cite{natdecouple} for dynamically vanishing when the scalar is in its ground state.

To see how $\varphi$ couples to matter localized on the branes, we generalize the previous discussion to include brane-localized matter fields, generically denoted by $\psi$. All brane quantities like tension and flux are regarded as being functions of both brane and bulk fields,
\be
 \tau_b = \tau_b(\psi, \phi) \,, \quad
 \Phi_b = \Phi_b(\psi, \phi)
 \quad \hbox{and so on.}
\ee
The main point is that none of this affects the matching conditions and solutions described above, and so in a static (or adiabatic) configuration the ground-state value $\varphi = \varphi_\star(\psi)$ still adjusts to satisfy
\be \label{equilibconfig}
 \sum_b \left[ \frac{\partial T_b(\psi, \phi)}{\partial \phi}
 \right]_{\varphi = \varphi_\star} = 0 \,.
\ee

The new ingredient that appears in searches for new forces is the use of spatially inhomogeneous matter configurations as sources ({\em e.g.} planets, stars, {\em etc.}) of spatial variation, $\delta \varphi = \delta \varphi(x)$, for the fluctuation $\delta \varphi = \varphi - \varphi_\star$ along the on-brane directions. Regarded graphically, these constrain the amplitude for emitting a single $\varphi$ particle from the source, with repeated emissions accumulating to give a coherent classical field. But the amplitude for $\varphi$-emission from matter localized on a specific brane, $b = b_0$, is controlled by the expansion of the brane action in powers of the fluctuation,
\be \label{forcecoupling}
 T_{b_0}(\psi, \varphi) = T_{b_0}(\psi, \varphi_\star) + \left[
 \frac{\partial T_{b_0}(\psi, \varphi)}{\partial \phi} \right]_{\varphi
 = \varphi_\star} \delta \varphi + \cO(\delta \varphi^2) \,.
\ee
Of these interactions, it is only the term linear in $\delta \varphi$ that acts as an obstruction to solving the field equations with $\delta \varphi = 0$, and so it is this linear term that is subject to the strongest constraint from new-force searches. Unless $T_{b_0}$ has special properties such a term could generate violations of the equivalence principle, which are strongly excluded once the range of the force becomes macroscopically large.

Comparing eqs.~\pref{equilibconfig} and \pref{forcecoupling} reveals the mechanism for suppressing $\varphi$-matter couplings. If the action, $T_{b_0}$, for the specific brane on which we live should share the same extremum as does the sum of all branes, $\sum_b T_b$, then as $\varphi_\star$ adjusts to satisfy the condition \pref{equilibconfig}, it would also automatically turn off the dangerous coupling of $\delta \varphi$ to matter localized at brane $b_0$. As the examples above show, the extremum of the sum of all brane actions need not agree with the extrema of each brane's action separately. But it automatically does so in two simple cases: $(i)$ when none of the branes besides $b_0$ couple to $\phi$ at all; and $(ii)$ when all of the branes couple to $\phi$, but are all extremal for the same place.

Notice that the argument is not changed by the presence of $\phi$-dependent brane fluxes, $\Phi_b$. This is because they do not enter into eq.~\pref{equilibconfig} independently from their contribution to $T_b$ (even though they do contribute independently to the value of $\vaceng$).

\section*{Acknowledgments}

We wish to thank Markus Luty and Raman Sundrum for useful discussions about back-reaction in codimension-two models. CB acknowledges the Kavli Institute for Theoretical Physics in Santa Barbara and the Abdus Salam International Center for Theoretical Physics for providing the very pleasant environs in which some of this work was performed, as well as Eyjafjallajokull for helping to provide some unexpected but undivided research time. LvN thanks the Instituut-Lorentz for Theoretical Physics at Leiden University for their hospitality. Our research is supported in part by funds from the Natural Sciences and Engineering Research Council (NSERC) of Canada. Research at the Perimeter Institute is supported in part by the Government of Canada through Industry Canada, and by the Province of Ontario through the Ministry of Research and Information (MRI).

\appendix

\section{Brane fluxes and flux quantization}
\label{App:fluxconditionws}

To see how to interpret the parameter $\Phi_b$, rewrite the brane flux term as a regularized 6D integral weighted by a scalar function $s(\rho)$ whose support is nonzero only in a short interval $|\rho - \rho_b| < \varepsilon$ away from the brane, and is normalized so that $\int \exd^2x \, \sqrt{g_2} \; s = 1$. That is,
\ba \label{app:6Dfluxform}
 S_\mathrm{flux} &=& \frac{\Phi_b}{2} \, \int \exd^6x \,\sqrt{- g_6} \; s \,
 \epsilon^{mn} \cF_{mn}  =   \Phi_b \, \int \exd^6x \, \sqrt{-g_4}
 \; s \, F_{\rho\theta} \,.
\ea
Then the $\delta \cA_\theta$ Maxwell equation becomes
\be
 \partial_\rho \Bigl( \sqrt{-g_6} \; F^{\rho\theta} - \Phi_b \sqrt{-g_4} \; s
 \Bigr) = 0 \,,
\ee
which integrates to give
\be \label{app:maxsolnws}
 e^{4W} \Bigl( e^{-B} \cA_\theta' - \Phi_b s \Bigr) = \cQ \,.
\ee
This is the bulk solution found in the text away from the brane, where $s = 0$.

Imagine now integrating this to obtain $\cA_\theta(\rho)$ in the vicinity of the brane at $\rho_b = 0$, using for $s$ a simple step function: $s = 1/(\pi \varepsilon^2)$ for $\rho < \varepsilon$ and $s = 0$ for $\rho > \varepsilon$. Assuming $W \simeq W_b$ is approximately constant and $e^B \simeq \rho$ for $\rho < \varepsilon$, the solution satisfying $\cA_\theta(0) = 0$ is
\be
 \cA_\theta(\rho) = \frac12 \, \left( \cQ e^{-4W_b} + \frac{\Phi_b}{\pi \varepsilon^2}
 \right) \rho^2 \,,
\ee
and so at $\rho = \varepsilon$ in particular
\be
 \cA_\theta(\varepsilon) = \frac{\Phi_b}{2 \pi} + \cO(\varepsilon^2) \,.
\ee

The junction condition for $\cA_\theta'$ at $\rho = \varepsilon$ can also be seen by subtracting the solution, eq.~\pref{app:maxsolnws} evaluated at $\rho < \varepsilon$ --- where $s = 1/(\pi \varepsilon^2)$ --- from the same solution evaluated at $\rho > \varepsilon$ --- where $s = 0$. Since the RHS is the same in both cases we get the following jump discontinuity across $\rho = \varepsilon$:
\be
 \Bigl[ e^{-B} \cA_\theta' \Bigr]^{\rho = \varepsilon+}_{\rho = \varepsilon-}
 = - \frac{\Phi_b}{ \pi \varepsilon^2} \,.
\ee
This can be related to the derivative of the brane action with respect to $\cA_\theta$ by rewriting eq.~\pref{app:6Dfluxform} as
\be
 S_\mathrm{flux} = \Phi_b \, \int \exd^6x \, \sqrt{-g_4}
 \; s \, F_{\rho\theta}  = \frac{2\pi \Phi_b}{\pi \varepsilon^2} \,
 \int \exd^4x \, \sqrt{- g_4} \; \cA_\theta(\varepsilon)
 \,,
\ee
and so (keeping in mind the relative sign between the tension and flux terms)
\be
 \Bigl[ e^{-B} \cA_\theta' \Bigr]^{\rho = \varepsilon+}_{\rho = \varepsilon-}
 = + \frac{1}{2 \pi} \; \left( \frac{\partial T_b}{\partial \cA_\theta}
 \right) \,,
\ee
as stated in ref.~\cite{BBvN}.

\section{Rugby-ball response}
\label{App:rugbyballresponse}

This section provides the explicit solutions for the properties of rugby ball solutions as functions of the assumed (shared) brane tension, and in particular computes the response to small changes in its value.

Rugby ball configurations solve the field equations
\be \label{app:EinsteineqNS}
 \cR_{\ssM\ssN} + \partial_\ssM \phi \, \partial_\ssN \phi
 + \kappa^2  \cF_{\ssM \ssP} {\cF_\ssN}^\ssP
  - \left[ \frac{\kappa^2}{8} \,
 \cF_{\ssP\ssQ} \cF^{\ssP \ssQ}
 - \frac{\kappa^2 \Lambda}{2}  \right]
 g_{\ssM\ssN} = 0 \,,
\ee
and
\be \label{app:ssmaxwelleqNS}
 \nabla_\ssM \cF^{\ssM \ssN} = 0 \,,
\ee
subject to the ansatz
\ba \label{app:bulksoln1}
 \exd s^2 &=& \hat g_{\mu\nu} \, \exd x^\mu \exd x^\nu
 + \exd \rho^2 + \alpha^2 L^2 \sin^2\left(
 \frac{\rho}{L}\right) \exd \theta^2 \\
 \cF_{\rho\theta} &=& \alpha \cQ L \sin\left(
 \frac{\rho}{L}\right), \label{bulksoln2}
\ea
with $\phi = \varphi_0$ constant. The bulk field equations give the 2D and 4D curvature scalars as
\begin{equation}
 \label{app:rugby constraints}
 -\cR_{(2)} = \frac{2}{L^2} = \kappa^2 \left(
  \frac{3\cQ^2}{2} + \Lambda \right) \,,
\end{equation}
and
\begin{equation} \label{app:4Dcurvature}
 \cR_{(4)} = {\hat \cR} = {2\kappa^2} \left( \frac{\cQ^2}{2}-\Lambda \right).
\end{equation}

The gauge potential corresponding to eq.~\pref{bulksoln2} is
\be
 \cA^\pm = \alpha \cQ L^2 \left[ \pm 1  - \cos \left(
 \frac{\rho}{L} \right) \right] \, \exd \theta\, \pm\frac{\Phi_\pm}{2\pi} \, \exd \theta \,,
\ee
where the $\pm$ sign indicates the solution for the northern or
southern hemisphere, since $\cA$ must evaluate to the brane flux at the corresponding pole. Requiring the difference between these two solutions near the equator to be a well-defined gauge transformation, $g A^+ - g A^- = \exd \Omega$, implies the constants $\cQ$ and $L$ must be related by
\be \label{app:quantizationcond}
 g \cQ = \frac{ N}{2 \, \alpha L^2} \,,
\ee
where we define $N=n-g\Phi_{\rm tot}/2\pi$.

Eqs.~\pref{app:rugby constraints} and \pref{app:quantizationcond}
determine the constants $\cQ$ and $L$ in terms of $\alpha$ and
$\Lambda$, with solutions
\be \label{app:Lsoln}
 \frac{1}{L^2} = \frac{8 \, \alpha^2 g^2 }{3 N^2 \kappa^2}
 \left[ 1 \pm \sqrt{ 1 - \left( \frac{3 \,N^2 \kappa^4 \Lambda}{
 8 \, \alpha^2 g^2} \right) } \right]
 = \frac{1 }{2 L_{\rm min}^2}
 \left[ 1 \pm \sqrt{ 1 - \left( \frac{3 \, N^2 \kappa^4 \Lambda}{
 8 \, \alpha^2 g^2} \right) } \right] \,,
\ee
and
\be \label{app:Bsoln}
 \cQ = \frac{N}{2 \alpha g L^2}
 = \frac{4 \alpha g}{3 N \kappa^2} \left[ 1 \pm
 \sqrt{1 - \left( \frac{3 \, N^2 \kappa^4 \Lambda}{
 8 \, \alpha^2 g^2} \right) } \right] \,.
\ee
These provide two solutions for $L$ and $\cQ$ for each given value
of $\alpha$ and $\Lambda$, satisfying $L^2 \ge L_{\rm min}^2 = 3
N^2 \kappa^2/16 \, \alpha^2 g^2$. Starting with the lower sign in
eq.~\pref{app:Lsoln} the radius $L$ falls from $L \to \infty$ to $L =
\sqrt 2 \; L_{\rm min}$ as $\Lambda$ climbs from 0 to
$\Lambda_{\rm max} = 8 \, \alpha^2 g^2/3 N^2 \kappa^4$. On this
branch $\Lambda \ll \Lambda_{\rm max}$ implies $1/L^2 \simeq
\kappa^2 \Lambda/2$. Then switching to the branch corresponding to
the upper sign has $L$ fall from $\sqrt 2\; L_{\rm min}$ to
$L_{\rm min}$ as $\Lambda$ recedes from $\Lambda_{\rm max}$ back
to zero. There are no real solutions with $\Lambda > \Lambda_{\rm
max}$, or with $L < L_{\rm min}$.

For each of these solutions the last equation,
eq.~\pref{app:4Dcurvature}, gives the on-brane curvature, $\hat \cR$.
There is a choice $\Lambda = \Lambda_f$, for which $\hat \cR$
vanishes, given by $\Lambda_f = \cQ^2/2$. For this choice $L$ and
$\cQ$ become
\be \label{app:L0B0soln}
 \frac{1}{L_f(\alpha)} = \frac{2 \, \alpha g}{N \kappa}
 \quad \hbox{and} \quad
 \cQ_f(\alpha) = \frac{2\,\alpha g}{N \kappa^2} \,,
\ee
and so
\be \label{app:Lambda0soln}
 \Lambda_f = \frac{\cQ^2_f}{2} = \frac{ 2\,
 \alpha^2 g^2}{N^2 \kappa^4} \,.
\ee
Because $L_{\rm min} < L_f < \sqrt 2 \; L_{\rm min}$ we see that
this solution lies on the branch corresponding to the upper sign
of eq.~\pref{app:Lsoln}. In particular
\be \label{app:nonsusyloident}
 \kappa^2 \cQ^2_f L_f^2 = \kappa^2 \left( \frac{2 \alpha
 g}{N \kappa^2} \right)^2 \left( \frac{N \kappa}{2 \alpha g}
 \right)^2 = 1 \,.
\ee

Notice that the semiclassical approximation requires the curvature
to remain small compared with the relevant energy scales, and so
in 6D requires $\cR^3$ to be much smaller than $\Lambda$ or $\cQ^2$.
Because $\cR \simeq \kappa^2 \Lambda$ and $\kappa^2 \cQ^2$ this
requires $\kappa^3 \Lambda$ and $\kappa^3 \cQ^2$ must both be much
smaller than unity. So for $\Lambda \sim \cQ^2 \sim \alpha^2
g^2/N^2 \kappa^4$ the semiclassical limit implies $\alpha^2
g^2/N^2 \kappa \ll 1$. This in turn ensures $\kappa/L^2_f \ll 1$,
showing that this value of $L_f$ lies within the classical limit.

The above expressions can be used to check the linearized analysis performed in the main text. To this end, suppose we start with $\alpha = \alpha_0$, with $\Lambda = \Lambda_0 = \Lambda_f(\alpha_0)$ chosen so that $\hat \cR = 0$ for this value of $\alpha$. Then we change the brane tension (but not the brane flux), and so also $\alpha$, without also adjusting $\Lambda$. Choosing the upper sign, the radius and magnetic flux become
\be \label{app:Lsoln0}
 \frac{1}{L^2} = \frac{8 \alpha^2 g^2 }{3 N^2 \kappa^2}
 \left[ 1 + \sqrt{ 1 - \left( \frac{3 N^2 \kappa^4 \Lambda_0}{
 8 \alpha^2 g^2} \right) } \right]
 = \left( \frac{2 \alpha^2}{3 \alpha_0^2} \right)
 \frac{1 }{L_0^2}
 \left[ 1 + \sqrt{ 1 -  \frac{3 \alpha^2_0}{
 4 \alpha^2}  } \right] \,,
\ee
and
\be \label{app:Bsoln0}
 \cQ = \frac{4 \alpha g}{3 N \kappa^2} \left[ 1 +
 \sqrt{1 - \left( \frac{3 N^2 \kappa^4 \Lambda_0}{
 8 \alpha^2 g^2} \right) } \right]
 = \left( \frac{2 \alpha}{3 \alpha_0} \right)
 \cQ_0 \left[ 1 + \sqrt{ 1 -
 \frac{3 \alpha^2_0}{
 4 \alpha^2} } \right] \,.
\ee
The last equality in these two equations is obtained by using
eq.~\pref{app:Lambda0soln} to trade $\Lambda_0$ for $\alpha_0$, and
then using eqs.~\pref{app:L0B0soln} to express the result in terms of
the values $L_0$ and $\cQ_0$ that correspond to $\alpha =
\alpha_0$. These equations show how the values of $L$ and $\cQ$
adjust to compensate for the change of $\alpha$. The on-brane
curvature similarly changes, and is given by
\ba
 \hat \cR &=& \kappa^2 (\cQ^2 - 2 \Lambda_0) \nn\\
 &=& \frac{32 \alpha_0^2 g^2}{9 N^2 \kappa^2} \left[ \sqrt{1
 - \frac{3 \alpha_0^2}{4\alpha^2}} + 1 -
 \frac{3 \alpha_0^2}{2\alpha^2} \right] \,,
\ea
which vanishes as $\alpha \to \alpha_0$, as it must. For $\alpha =
\alpha_0  + \Delta \alpha$, then $\alpha_0^2/\alpha^2 \simeq 1 - 2
\Delta \alpha/\alpha_0$ and so
\be
 \hat \cR \simeq \left( \frac{16 \alpha_0^2 g^2 }{ N^2 \kappa^2}
 \right) \left( \frac{\Delta \alpha}{\alpha_0} \right)
 = \frac{4}{L_0^2} \left( \frac{\Delta \alpha}{\alpha_0}
 \right)
 = -  \frac{2\kappa^2 \Delta T}{\pi \alpha_0 L_0^2}
 = -8 \kappa_4^2 \Delta T  \,,
\ee
which uses the matching condition $1-\alpha = 4GT = \kappa^2
T/2\pi$ in the form $\Delta \alpha = - \kappa^2 \Delta T/2 \pi$,
as well as the definition of the 4D gravitational coupling:
$\kappa^2 = 4 \pi \alpha_0 L_0^2 \, \kappa_4^2$. Defining the
4D potential, $V_{\rm eff}$, by $\hat \cR = -4 \kappa_4^2 \,
V_{\rm eff}$ gives the expected result
\be
 V_{\rm eff} \simeq 2
  \Delta T \,.
\ee
The factor of 2 arises because a change of $\alpha$ requires an
equal change of tension for {\em both} branes if it is to preserve the rugby-ball form.

\section{Misaligned currents}
\label{App:misalignedJ}

This Appendix uses a simple model to track the implications that arise if the external current happens not to be aligned precisely with the lightest mode of the system. When this happens errors can arise in the identification of quantities like low-energy masses, but this section argues that these are generically suppressed by powers of the light mass divided by heavier masses.

Consider then the toy 4D lagrangian
\be
 \frac{\cL}{\sqrt{-g}} = - \frac{1}{2 \kappa^2_4} \, \cR
 - \frac12 \Bigl[ (\partial \varphi)^2 + (\partial \chi)^2 \Bigr]
 - V(\varphi, \chi) \,,
\ee
whose potential is given by
\be
 V = V_0 + \frac12 \Bigl( m^2 \varphi^2 + M^2 \chi^2 \Bigr)
 + J (\varphi + \zeta \, \chi) \,,
\ee
with masses assumed to satisfy $m \ll M$. Here $\varphi$ is meant as the analog of the KK would-be zero mode in the main text, while $\chi$ is representative of some other, more massive, KK mode. The goal is to ascertain the extent to which our method of determining the low-energy mass would be thrown off by a small coupling --- parameterized here by $\zeta$ --- of the external current to a heavy state.

The classical equations of motion for the scalar fields are
\be
 \Box \, \varphi - m^2 \varphi = J
 \quad \hbox{and} \quad
 \Box \, \chi - M^2 \chi = \zeta \, J \,,
\ee
while the Einstein equation reads
\be
 \cR_{\mu\nu} + \kappa_4^2 \Bigl( \partial_\mu \varphi \, \partial_\nu \varphi
 + \partial_\mu \chi \, \partial_\nu \chi \Bigr)
 + \kappa_4^2 V = 0 \,,
\ee
and so
\be
 \frac{\cR}{4 \kappa_4^2} = -V = - V_0 - \frac12 \Bigl( m^2 \varphi^2 + M^2 \chi^2 \Bigr)
 - J (\varphi + \zeta \, \chi) \,.
\ee
Evaluated at the particular solutions
\be
 J = - m^2 \varphi
 \quad \hbox{and} \quad
 \chi = - \frac{\zeta J}{M^2}
 = \left( \frac{\zeta m^2}{M^2} \right) \varphi \,,
\ee
this last equation gives
\be
 \cF(\varphi) := \frac{\cR}{4 \kappa_4^2} = - V_0 + \frac12 \, m^2 \varphi^2
 \left[ 1 + \left( \frac{\zeta m}{M} \right)^2 \right] \,.
\ee

In terms of $\cF(\varphi)$ the method of the main text gives the low-energy scalar potential as
\be
 V_{\rm eff}(\varphi) := \varphi \int \frac{\exd \varphi}{\varphi^2} \;
 \cF(\varphi) \,.
\ee
For $\cF(\varphi) = A + B \varphi + \frac12 \, C \varphi^2$ the integral evaluates to\footnote{The singular form of $V_{\rm eff}''(0)$ when $B \ne 0$ corresponds to the pathological case where brane fluxes and tensions are not extremized for the same value of $\varphi$, discussed in more detail in Appendix \ref{App:branesvsflux}.}
\be
 V_{\rm eff} = -A + B \varphi \ln \varphi + \frac12 \, C \varphi^2 + D \varphi \,,
\ee
where $D$ is the integration constant, and so when applied to the above toy model this gives
\be
 V_{\rm eff} (\varphi) = V_0 + \frac12 \, m^2 \varphi^2
 \left[ 1 + \left( \frac{\zeta m}{M} \right)^2 \right] \,.
\ee
This expression correctly identifies the value of the potential at its minimum to be $V_0$, and --- provided $\zeta \lsim \cO(1)$ --- gives the correct mass for the field $\varphi$, up to corrections of relative order $m^2/M^2$.

\section{When brane fluxes and tensions compete}
\label{App:branesvsflux}

This Appendix briefly discusses another kind of competition, which would arise if $\delta T_b (\varphi_0)$ and $\delta\Phi_b (\varphi_0)$ at the same brane were not minimized by the same scalar configuration. A simple representative in this category is
\be
 T_\ssN (\varphi_0) = T+ T_{\ssN0} + \frac{T_{\ssN2}}{2} \, (\varphi_0 - \pphi_\ssT)^2 \quad \hbox{and} \quad
 \Phi_\ssN (\varphi_0) = \Phi + \Phi_{\ssN0} + \frac{\Phi_{\ssN2}}{2} \, (\varphi_0 - \pphi_\phi)^2 \,,
\ee
together with $\delta T_\ssS (\varphi_0) = \delta \Phi_\ssS (\varphi_0) = 0$, so the `south' brane plays no role in the stabilization of $\varphi_0$.

In this case because the flux is irrelevant for determining $\varphi_\star$, its value is simply $\varphi_\star = \pphi_\ssT$. The warping difference is also insensitive to $\Phi_b$ and so becomes $W_\ssN - W_\ssS = (3 \kappa^2 T_{\ssN0}/20 \pi \alpha)$, while $\delta L/L = (3 \kappa^2 \vaceng/8\pi \alpha)$ with
\ba
 \vaceng &=& T_{\ssN0}  - \cQ \Phi_{\ssN 0}
  - \frac{ \cQ \Phi_{\ssN2}}{2} \, (\pphi_\ssT - \pphi_\phi)^2  \nn\\
 \hbox{and} \quad
 m_\varphi^2 &=& \frac{T_{\ssN 2}}{f^2} +
 \frac{\cQ \Phi_{\ssN2}}{f^2} \lim_{\varphi\rightarrow\pphi_\ssT} \left( \frac{\varphi - \pphi_\phi}{\varphi-\pphi_\ssT}
 \right)  \,.
\ea
Clearly, the expression for the mass is singular when $\pphi_\ssT\neq\pphi_\phi$. The reason for the singularity, is that for this choice of brane there is no solution satisfying the \emph{ansatz} with which we work. The obstruction lies with the Maxwell field, which we choose to lie in the $\cF_{\rho\theta}$ direction only. However, the perturbation that gets excited by moving $\varphi_0$ away from equilibrium, if we do not stabilize with an external current, necessarily gets a time dependent Maxwell field. But a changing magnetic field induces an electric field, so the $\cF_{\rho t}$ and $\cF_{\theta t}$ components cannot both remain zero.

To see that the Maxwell field must acquire a time dependence, consider a perturbation, $\delta \varphi$, that oscillates about the background $\pphi_\ssT$,
\be
 \varphi = \pphi_\ssT + \delta \varphi(\rho) \, e^{-imt} \,.
\ee
In the flux condition, eq.~\pref{fluxquantznlin}, the brane fluxes now have a part that is linear in $\delta \varphi$ that becomes proportional to $e^{-imt}$. If we now assume that $\delta\cQ$ does not acquire any time dependence, we get a contradiction: $\delta B$ has a part proportional to $e^{-imt}$ according to the flux condition, but in matching it with the brane in eq.~\pref{Bwithcurrent}, if $\delta \cQ$ is time independent the right hand side is either constant or proportional to $(\Phi - \pphi_\ssT)^2 \propto e^{-2imt}$. This shows that the matching conditions cannot be satisfied unless the magnetic field becomes time-dependent.

\end{document}